# 4MOST Cosmology Redshift Survey (CRS): clustering properties of CRS bright galaxy and luminous red galaxy target catalogues

Behnood Bandi ,[1★] Antoine Rocher,[2] Aurélien Verdier ,[2] Jon Loveday ,[1] Zhuo Chen,[2] Johan Richard ,[3] Jean-Paul Kneib,[2] Tom Shanks [4] and Michael J. I. Brown [5]

[1]*Astronomy Centre, Department of Physics and Astronomy, University of Sussex, Falmer, Brighton BN1 9QH, UK*
[2]*Laboratoire d'astrophysique, Ecole Polytechnique Fédérale de Lausanne (EPFL) Observatoire, CH-1290 Versoix, Switzerland*
[3]*Univ Lyon, Univ Lyon1, ENS de Lyon, CNRS, Centre de Recherche Astrophysique de Lyon UMR5574, Saint-Genis-Laval, France*
[4]*Department of Physics, Centre for Extragalactic Astronomy, Durham University, South Road, Durham DH1 3LE, UK*
[5]*School of Physics & Astronomy, Monash University, Clayton, VIC 3800, Australia*

Accepted 2026 February 25. Received 2026 February 25; in original form 2025 October 2

## ABSTRACT

The 4-m Multi-Object Spectroscopic Telescope Cosmology Redshift Survey (4MOST CRS) will obtain 5.4 million spectroscopic redshifts over $\sim 5700\,\mathrm{deg}^2$ to map large-scale structure and enable measurements of baryon acoustic oscillations, growth rates via redshift-space distortions, and cross-correlations with weak-lensing surveys. We validate the target selections, photometry, masking, systematics, and redshift distributions of the bright galaxy (BG) and luminous red galaxy (LRG) target catalogues selected from the Dark Energy Spectroscopic Instrument (DESI) Legacy Survey DR10.1 (LS) imaging. We measure the angular two-point correlation function, $w(\theta)$, test masking strategies, and recover redshift distributions via cross-correlation with DESI DR1 spectroscopy. For BGs, we adopt LS `MASKBITS` that veto bright stars and extended sources; for LRGs, we pair these with unblurred coadds of the WISE imaging (unWISE) W1 artefact masks. These choices suppress small-scale excess power without imprinting large-scale modes. A Limber-scaling test across BG $r$-band magnitude slices shows that, after applying the scaling, the $w(\theta)$ curves collapse to a near-common power law, demonstrating photometric uniformity with depth and consistency between the North and South Galactic Caps. Cross-correlations with DESI spectroscopy recover the expected $N(z)$, albeit with high shot noise at the brightest magnitudes. For LRGs, angular clustering in photo-$z$ slices ($0.4 \leq z < 1.0$) is consistent between the Dark Energy Camera Legacy Survey (DECaLS) and Dark Energy Survey footprints and is well described by an approximate power law once photo-$z$ smearing is included; halo-occupation fits are consistent with recent LRG studies. Together, these tests indicate that the masks and target selections yield reliable clustering statistics, supporting precision large-scale structure analyses with 4MOST CRS.

## 1 INTRODUCTION

The 4-m Multi-Object Spectroscopic Telescope (4MOST; R. S. Jong et al. 2019) is a powerful new facility installed on the Visible and Infrared Survey Telescope for Astronomy (VISTA) at Paranal Observatory. With a $4.2\,\mathrm{deg}^2$ field of view and the capacity to observe more than 2400 targets simultaneously, 4MOST is designed to carry out a diverse suite of spectroscopic surveys targeting over $2 \times 10^7$ astrophysical sources over a 5-yr operational period. Its high multiplexing capability and broad sky coverage enable it to address major scientific themes, including galaxy formation, Galactic archaeology, and cosmology.

4MOST will also play a crucial role in complementing major international projects such as the Rubin Observatory's Legacy Survey of Space and Time (LSST; Ž. Ivezić et al. 2019), *Euclid* (Euclid Collaboration 2025), and the Square Kilometre Array (SKA; P. Dewdney et al. 2009), with a particular emphasis on the southern sky. The spectroscopic data collected by 4MOST will enable precise measurements of redshifts, stellar parameters, and elemental abundances, improving our understanding of galaxy assembly histories, the structure of the Milky Way, and the nature of dark matter and dark energy. In addition, 4MOST's design allows multiple science programmes to run in parallel, optimising observational efficiency and delivering broad, uniform data sets (R. S. Jong et al. 2019).

A key survey of 4MOST is the Cosmology Redshift Survey (CRS; J. Richard et al. 2019), designed to trace the growth of large-scale structure and to constrain the nature of dark energy and gravity on cosmological scales. CRS will obtain spectroscopic red-shifts for approximately 5.4 million galaxies and quasars over $\sim$ 5700 deg$^2$ in the southern hemisphere, spanning $z \simeq 0.15$ to 3.5. Through measurements of galaxy clustering, redshift-space distortions (RSDs), and cross-correlations with weak lensing, CRS aims to deliver precise constraints on cosmic acceleration and the growth rate of structure.

★ E-mail: b.bandi@sussex.ac.uk

A particular strength of CRS is its extensive overlap with leading southern imaging surveys, including the Dark Energy Survey (DES), the Kilo-Degree Survey (KiDS; A. H. Wright et al. 2024), and, most importantly, the LSST. This synergy enables powerful cross-correlation analyses that combine CRS spectroscopy with weak gravitational lensing and other complementary data sets. In particular, CRS can help to calibrate photometric redshifts and control galaxy-bias systematics, thereby sharpening cosmological constraints. As a result, 4MOST CRS is well placed to make significant contributions to tests of gravitational physics and to the determination of cosmological parameters.

Realising these goals requires a well-understood and spatially uniform spectroscopic target selection. The CRS target selection comprises three sub-surveys tailored to different redshift ranges and populations, notably bright galaxies (BGs), luminous red galaxies (LRGs), and quasars. While earlier versions of the BG and LRG selections were based on VISTA photometry (J. Richard et al. 2019), the latest catalogues adopt DESI Legacy Surveys DR10.1 imaging, benefiting from deeper data, more homogeneous calibration, and improved masking (A. Verdier et al. 2025, VR25). These updates also align CRS with DESI-like target definitions, facilitating direct comparisons and joint analyses.

In this paper, we present a clustering-based validation of the 4MOST CRS BG and LRG target catalogues selected from Legacy Survey DR10.1 photometry. In Section 2, we describe the target selection, masking, and data sets used in this analysis. Section 3 presents measurements of the angular two-point correlation function, $w(\theta)$, used to assess sample quality and the impact of masking strategies. In Section 4, we apply Limber's equation to test the consistency of angular clustering across BG magnitude slices, and we validate the implied real-space clustering using projected correlation measurements, $w_p(r_p)$, from DESI DR1. Section 5 introduces clustering-redshift (cluster-$z$) techniques to estimate the redshift distributions of BG targets. Section 6 measures the angular clustering of LRGs in photometric-redshift slices across DECaLS and DES regions, and Section 7 models the LRG projected clustering with a halo-occupation framework. The section summarises our findings and discusses implications for early CRS cosmology.

Throughout the paper, unless stated otherwise, we adopt a flat Lambda cold dark matter ($\Lambda$CDM) cosmology consistent with the *Planck* 2018 parameters (Planck Collaboration VI 2020).

## 2 DATA AND CRS TARGET CATALOGUES

Initially, the target selection of the BGs and LRGs, which was introduced in J. Richard et al. (2019), was based on the VISTA Hemisphere Survey (VHS), VISTA Kilo-Degree Infrared Galaxy Survey (VIKING; A. Edge et al. 2013), and Wide-field Infrared Survey Explorer (WISE; E. L. Wright et al. 2010) photometry. However, the target selection of these two sub-surveys has been changed, and the current version, described in VR25, uses the DESI Legacy Survey DR10.1 (LS 10.1). This decision was motivated by Legacy Survey's superior photometric quality and depth and its compatibility with Dark Energy Spectroscopic Instrument (DESI; DESI Collaboration et al. 2025) data, facilitating cross-survey analyses and cosmological parameter constraints.

LS 10.1is a combination of three individual surveys. The Mayall $z$-band Legacy Survey (MzLS; A. Dey et al. 2016) and the Beijing-Arizona Sky Survey (BASS; H. Zou et al. 2017) imaged the northern sky above $\delta \geq 32^\circ$. The Dark Energy Camera Legacy Survey (DECaLS; B. Flaugher et al. 2015) provided the optical imaging of the sky that covers both the North Galactic Cap region at $\delta \leq 32^\circ$ and the South Galactic Cap region at $\delta \leq 34^\circ$. In addition, the photometry of the DES using the same camera as DECaLS is used to cover the southern hemisphere with higher depth.

In this section, we briefly introduce the BG and LRG target selections using the LS 10.1and discuss the differences between the original and current target selections. Also, we describe other data products that we have used in this work.

### 2.1 CRS target selection from legacy survey

#### 2.1.1 CRS-BG photometric catalogue

The BG targets of the 4MOST-CRS are selected from LS 10.1, supplemented by astrometric and photometric data from *Gaia* Early Data Release 3 (EDR3) and the Tycho-2 bright star catalogue. The target selection criteria for BG galaxies are adapted from the DESI Bright Galaxy (BGS) (C. Hahn et al. 2023) with additional selections tailored to meet the specific observational constraints and science goals of the 4MOST project (VR25). These extra selections consist of cuts to remove stars using *Gaia* proper motion information, an additional colour selection to remove low redshift targets, and reduction of the magnitude limit to $r_{mag} < 19.25$ to reduce the target density to $\sim 250\,\mathrm{deg}^{-2}$ (compared to $\sim 860\,\mathrm{deg}^{-2}$ in DESI). The colour selection to remove low redshift targets is detailed in section 3.2.1 of VR25. All maskings that we used for the BG catalogue are from Legacy Survey BITMASKS.[1] `MASKBIT 1` is used to mask objects around *Gaia* and Tycho-2 bright stars, and sources around *Gaia* stars with $G < 16$ are masked using `MASKBIT 11`. `MASKBIT 12` and `BITMASK 13` are used to remove areas around large galaxies and globular clusters, respectively. Section 3.2 studies the effect of masking on the angular correlation functions. Table 1 describes the masks used for both the BG and LRG sub-surveys. The radii for both `MASKBIT 1` and `MASKBIT 11` are defined as functions of the *Gaia* $G$-band magnitude (described in LS documentation). `MASKBIT 1` is referred to as the 'bright-star' mask because it is applied only to bright *Gaia* and Tycho-2 stars ($G < 13$). The `MASKBIT 11` radius is $R_{\mathrm{deg}} = (1630/3600)\,1.369^{-G_{\mathrm{Gaia}}}$.

#### 2.1.2 CRS-LRG photometric catalogue

The LRG target catalogue aims to efficiently sample galaxies within an intermediate redshift range, specifically targeting $0.4 < z < 1$. The selection criteria are broadly derived from the DESI LRG sample (R. Zhou et al. 2023), but similar to the BG targets, the DESI LRG selection has been modified to reduce the target density of the catalogue to $\sim 400\,\mathrm{deg}^{-2}$ (compared to $\sim 600\,\mathrm{deg}^{-2}$ in DESI). These modifications remove fainter targets at higher redshifts and lower the target density (VR25). Initially, just `MASKBITs` 1, 12, and 13 were used to mask the LRG sample, but after performing angular clustering tests, described in Section

[1] https://www.legacysurvey.org/dr10/bitmasks/

**Table 1.** Description of LS `MASKBITS` and `WISEMASK_W1` flags used in the BG and LRG target catalogues. Based on the LS bitmask documentation.

| Bit | Name | Description | Subsurveys |
|---|---|---|---|
| | | **LSMASKBITS** | |
| 1 | BRIGHT | Bright stars: MAG_VT $<$ 13 (Tycho-2) or $G$ $<$ 13 (*Gaia*) | BG and LRG |
| 11 | MEDIUM | *Gaia* stars with $G$ $<$ 16 | BG and LRG |
| 12 | GALAXY | Large galaxies from SGA | BG and LRG |
| 13 | CLUSTER | Globular clusters | BG and LRG |
| | | **WISEMASK_W1** | |
| 0 | BRIGHT | Bright star core and wings | LRG |
| 1 | SPIKE | PSF-based diffraction spike | LRG |
| 2 | GHOST | Optical ghost | LRG |
| 3 | LATENT | First latent image | LRG |
| 4 | LATENT2 | Second latent image | LRG |
| 5 | HALO | AllWISE-like circular halo | LRG |
| 6 | SATUR | Bright star saturation | LRG |
| 7 | SPIKE2 | Geometric diffraction spike | LRG |

3.2, we added `MASKBIT` 11 and all `WISEMASK_W1` bitmasks, for unWISE W1 band maskings.[2]

## 2.2 DESI DR1

We utilise the large-scale structure (LSS) catalogues from the first DESI Data Release (A. J. Ross et al. 2024; DESI Collaboration et al. 2025) for multiple purposes: (i) to estimate the cumulative redshift distribution, (ii) to model the spatial two-point correlation function, and (iii) to cross-correlate with photometric samples in order to derive clustering redshift (cluster-z) estimates.

To ensure consistency with the 4MOST CRS selection, we apply the CRS target selection criteria to the DESI DR1 data. This allows us to construct a subset of DESI with comparable redshift and magnitude distributions, enabling direct comparisons between DESI and CRS samples in subsequent analyses.

## 2.3 Random catalogues

For angular clustering, we use the Legacy Surveys imaging randoms at a density of 2500 deg$^{-2}$ (A. D. Myers et al. 2023). We use these randoms only to trace the angular selection: the imaged footprint and the veto masks. Accordingly, we require NOBS $>$ 0 so that points lie within the imaged area, and we apply the relevant `MASKBITS` listed in Table 1. We combine enough random files that the total number of random points is more than ten times the number of targets, which stabilises pair counts and uncertainties. We do not attempt to imprint the photometric target selection on the randoms. See A. D. Myers et al. (2023) for a description of the random catalogue generation.

For the projected correlation function $w_p(r_p)$ based on DESI DR1, we use the DESI DR1 random catalogues, which include redshifts. We apply the same tracer definition and redshift cuts as for the data, together with the maskbits in Table 1. The redshifts in the DESI randoms allow consistent pair counts in $(r_p, \pi)$ and the projection to $w_p(r_p)$.

# 3 ANGULAR CLUSTERING MEASUREMENTS

## 3.1 Two-point correlation function and angular clustering

The two-point correlation function, often denoted as $\xi(r)$, is a statistical tool used to quantify the clustering of galaxies in the Universe. It measures the excess probability of finding a pair of galaxies separated by distance ($r$), compared to what would be expected in a random distribution and can be defined as

$$\mathrm{d}P = \bar{n}^2 \left[1 + \xi(r)\right] \mathrm{d}V_1 \mathrm{d}V_2, \tag{1}$$

where $\bar{n}$ is the mean comoving number density of galaxies, and $dV_1$, $dV_2$ are differential volume elements (Peebles 1980).

The angular two-point correlation function, denoted as $w(\theta)$, measures the excess probability of finding two galaxies separated by an angle, $\theta$, on the sky. It is a projection of the three-dimensional clustering on to the two-dimensional celestial sphere (A. L. Coil 2012). The angular correlation function is particularly useful for studying the clustering of galaxies at high redshifts or for faint galaxy populations, where obtaining spectroscopic redshifts may be challenging. By analysing the angular correlation function, we can still infer valuable information about the large-scale structure of the Universe and the properties of galaxies, even without precise distance measurements. Over a wide range of angles, $w(\theta)$ follows a power law defined as (E. J. Groth & P. J. E. Peebles 1977):

$$w(\theta) = A_w \theta^{1-\gamma}, \tag{2}$$

where $A_w$ is the clustering amplitude at a given scale and $1 - \gamma$ is the power-law index.

Computing the two-point correlation function, $\xi(r)$ or $w(\theta)$, involves counting pairs of galaxies according to their separation and dividing this by what we expect from an unclustered distribution. To perform pair counting, we should create a random catalogue with identical sky coverage to our data but with points randomly distributed. We measure the angular correlation function $w(\theta)$ by comparing galaxy–galaxy pair counts in the data with those from a matched random catalogue. Among the available estimators for $w(\theta)$, we adopt the Landy–Szalay estimator (S. D. Landy & A. S. Szalay 1993), which is widely used in clustering analyses.

The Landy–Szalay estimator for calculating autocorrelation functions is

$$w(\theta) = \frac{1}{RR}\left[DD - 2DR + RR\right], \tag{3}$$

where RR, DR, and DD are pair counts for the random-random, data-random, and data-data catalogues, respectively, and the general form of it for cross-correlation (used in Section 5) can be written as

$$w(\theta) = \frac{1}{RR}\left[D_1D_2 - D_1R - RD_2 + RR\right]. \tag{4}$$

The angular correlation functions ($w(\theta)$) have been computed using `the TreeCorr` python package (M. Jarvis, G. Bernstein & B. Jain 2004). projected correlation functions ($w_p(r_p)$) and angular correlation functions of the LRG catalogue in redshift slices were estimated using `pycorr`,[3] which is a wrapper for `Cor-`

[2]Aaron Meisner's unWISE masks documentation https://catalog.unwise.me/files/unwise_bitmask_writeup-03Dec2018.pdf

[3]https://github.com/cosmodesi/pycorr

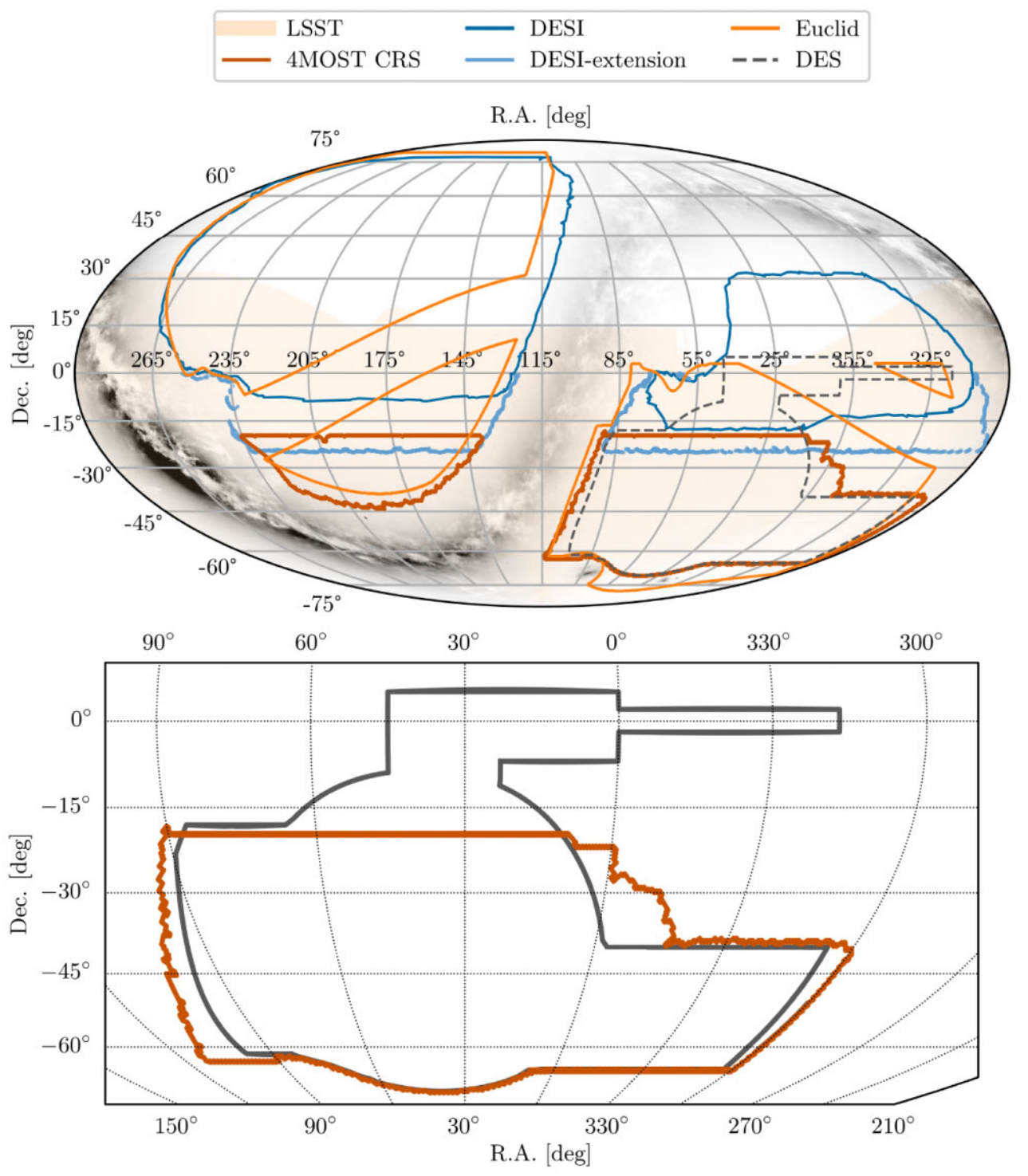

**Figure 1.** *Top panel:* footprint of CRS-BG and CRS-LRG as well as other surveys. *Bottom panel:* footprint of the 4MOST-CRS SGC sky and the DES photometric survey. The regions of 4MOST-CRS outside DES used DECaLS photometric data. This illustrates that the SGC footprint CRS is mostly using DES photometric data.

rFunc (M. Sinha & L. Garrison 2019b; M. Sinha & L. H. Garrison 2020). To estimate errors on the data measurements, we used the Jackknife (JK) method (C.-F. J. Wu 1986) with 36 independent regions. The jackknife regions were defined using a K-means sampler that cuts the footprint into regions of similar size in RA/DEC, as implemented in the DESI package pycorr.

The photometry of both LRG and BG samples is based on DECaLS and DES, which have different depths that could affect target selection. The CRS footprint in NGC is based on DECaLS, while the SGC is mostly in the DES region, with small areas in DECaLS, as illustrated in Fig. 1. To test the influence of the different photometric regions, we perform a clustering analysis independently within the two survey regions throughout the paper.

### 3.2 Mask evaluation using angular correlation function

Accurate measurements of the angular two-point correlation function, $w(\theta)$, require careful control of observational systematics. Artefacts, bright stars, and large nearby sources can induce spurious pairs or mask real ones, biasing the estimate. We therefore tested several combinations of Legacy Surveys DR10.1 (LS) MASKBITS and unWISE artefact masks for the 4MOST CRS BG and LRG catalogues, and measured $w(\theta)$ for each configuration.

Fig. 2 summarises the results. The *upper* panels show $w(\theta)$ for each masking choice. For LRG (right), progressively stricter masking decreases the small-scale amplitude and yields a cleaner power-law behaviour, consistent with removing false LRG detections around bright stars and WISE artefacts. For BG (left), both amplitude and slope vary only mildly across masks, indicating weaker sensitivity at the tested depths; the small-scale points are nevertheless most stable when bright-star and extended-source masks are applied.

The lower panels test the stability of the measurement by comparing each masking configuration to the final adopted mask:

$$\Delta w(\theta) \equiv w_{\rm final}(\theta) - w_{\rm mask}(\theta). \tag{5}$$

The uncertainty is estimated using a joint jackknife approach, where the same sky regions are removed from both measurements. For each jackknife realisation$r$, we compute

$$\Delta w^{(r)}(\theta) = w^{(r)}_{\rm final}(\theta) - w^{(r)}_{\rm mask}(\theta), \tag{6}$$

and calculate the mean difference

$$\overline{\Delta w}(\theta) = \frac{1}{N}\sum_{r=1}^{N} \Delta w^{(r)}(\theta), \tag{7}$$

where $N$ is the total number of jackknife regions. The covariance matrix of $\Delta w(\theta)$ is then

$$C^{(\Delta w)}_{ij} = \frac{N-1}{N}\sum_{r=1}^{N}\Big[\Delta w^{(r)}(\theta_i) - \overline{\Delta w}(\theta_i)\Big]\Big[\Delta w^{(r)}(\theta_j) - \overline{\Delta w}(\theta_j)\Big], \tag{8}$$

and the $1\sigma$ uncertainty in each angular bin is

$$\sigma(\theta_i) = \sqrt{C^{(\Delta w)}_{ii}}. \tag{9}$$

This method accounts for the correlation between the two measurements and provides smaller, more realistic uncertainties than treating them as independent. Jackknife errors were calculated using the built-in tools of TreeCorr.

We display the normalised difference $\Delta w(\theta)/\sigma(\theta)$ to show how each successive masking step affects the correlation function. As masking becomes more restrictive, objects affected by bright stars or extended sources are removed, and the measurement should converge. We define stability as the regime where further masking leads to changes $|\Delta w(\theta)| \lesssim \sigma(\theta)$ across most scales. Small fluctuations are expected because of sample variance and angular covariance, but once the distribution of $\Delta w(\theta)/\sigma(\theta)$ narrows around zero, the measurement can be considered stable.

In the LRG panels (right), looser masks generally yield negative $\Delta w(\theta)/\sigma(\theta)$ at small angles, i.e. $w_{\rm final} < w_{\rm mask}$. This is expected if stricter masks remove diffraction-spike and halo detections misclassified as LRGs, which otherwise contaminate the clustering signal. Using the LS MASKBITS (11, 12, and 13; bright stars, large nearby galaxies, globular clusters) alone does not reach the stability regime: residual deviations remain at small scales relative to the final mask. Adding unWISE artefact masks suppresses these residuals and brings $\Delta w(\theta)/\sigma(\theta)$ close to zero across most $\theta$, so we adopt both LS and unWISE MASKBITS.

In the BG panels (left), $\Delta w(\theta)/\sigma(\theta)$ remains close to zero across most scales and masking configurations, consistent with the modest variations seen in the upper panels. Nevertheless, the 2D histograms of the position of targets around *Gaia* stars in Fig. 3 show clear overdensities around *Gaia* stars with $7 < G < 11$ and still present for $13 \leq G < 16$, which supports adopting medium-star masks (MASKBIT 11) that extend beyond the MASKBIT 1 radii and masks stars with $G < 16$. We therefore use LS MASKBITS 11, 12, and 13 for BG; with these applied, additional masking changes produce shifts that are statistically insignificant over the angular range considered. Fig. 4 shows two cut-outs from the Legacy Surveys Sky Viewer[4] around two randomly selected stars. Horizontal streaks in Figs 3 and 4 through saturated stars are CCD bleed trails (blooming), formed when charge spills from saturated pixels and is transported along the detector read-out direction. After resampling to the north-up RA–Dec. grid used by the Legacy Surveys, these trails appear as near-constant-declination lines, i.e. aligned with the RA axis, and should not be confused with diffraction spikes (F. Valdes, R. Gruendl & DES Project 2014; A. Dey et al. 2019). These artefacts are captured by `MASKBITS` 5–7 (the per-band `ALLMASK_g`, `ALLMASK_r`, `ALLMASK_z` flags) and can be removed by applying these masks. We do not impose these cuts at the target-selection stage; instead, we can filter the affected regions during downstream catalogue post-processing. In Figs 3 and 4, $R_{\mathrm{BS}}$ denotes the `MASKBIT 1` radius.

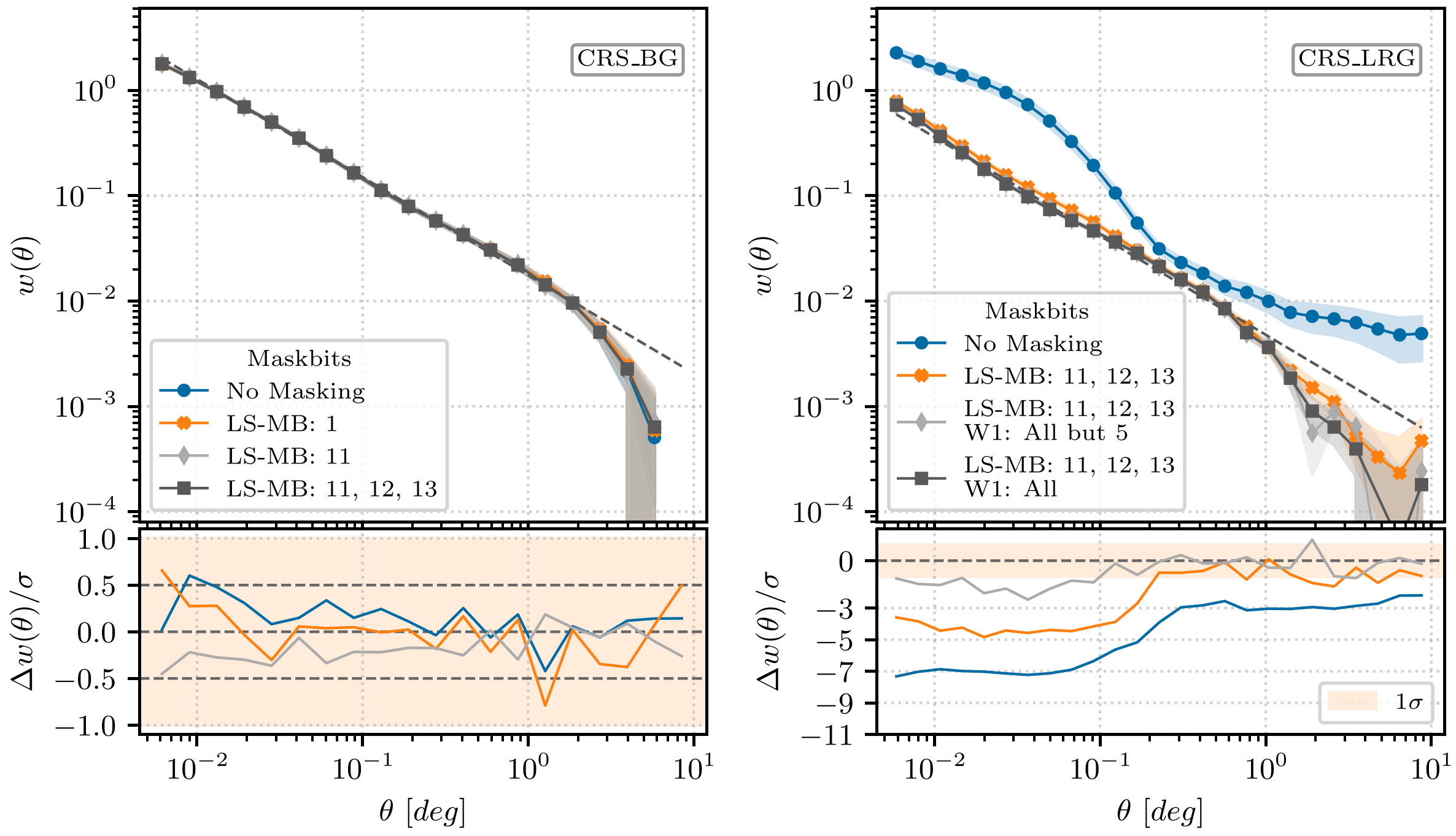

**Figure 2.** Angular correlation functions for different masking choices. *Left panels*: BG; *right panels*: LRG. Upper panels: $w(\theta)$ with uncertainties from 36 jackknife regions. Lower panels: differences relative to the adopted final masking, shown as $\Delta w(\theta)/\sigma(\theta)$ where $\Delta w(\theta) \equiv w_{\rm final}(\theta) - w_{\rm mask}(\theta)$ and $\sigma(\theta)$ is the jackknife estimate of error in the difference (equations 5 to 9). The negative values indicate that the final mask decreases the measured clustering amplitude compared to a given configuration, as expected when stricter masks remove contaminants. The LRG panels show a clear monotonic approach towards zero with stricter masking, consistent with the removal of star-spike and unWISE artefact detections misclassified as LRGs. BG clustering is largely stable but still benefits from robust bright-star masking. The dashed lines in the upper panels show power-law fits (equation 2) for the final masking used for each tracer.

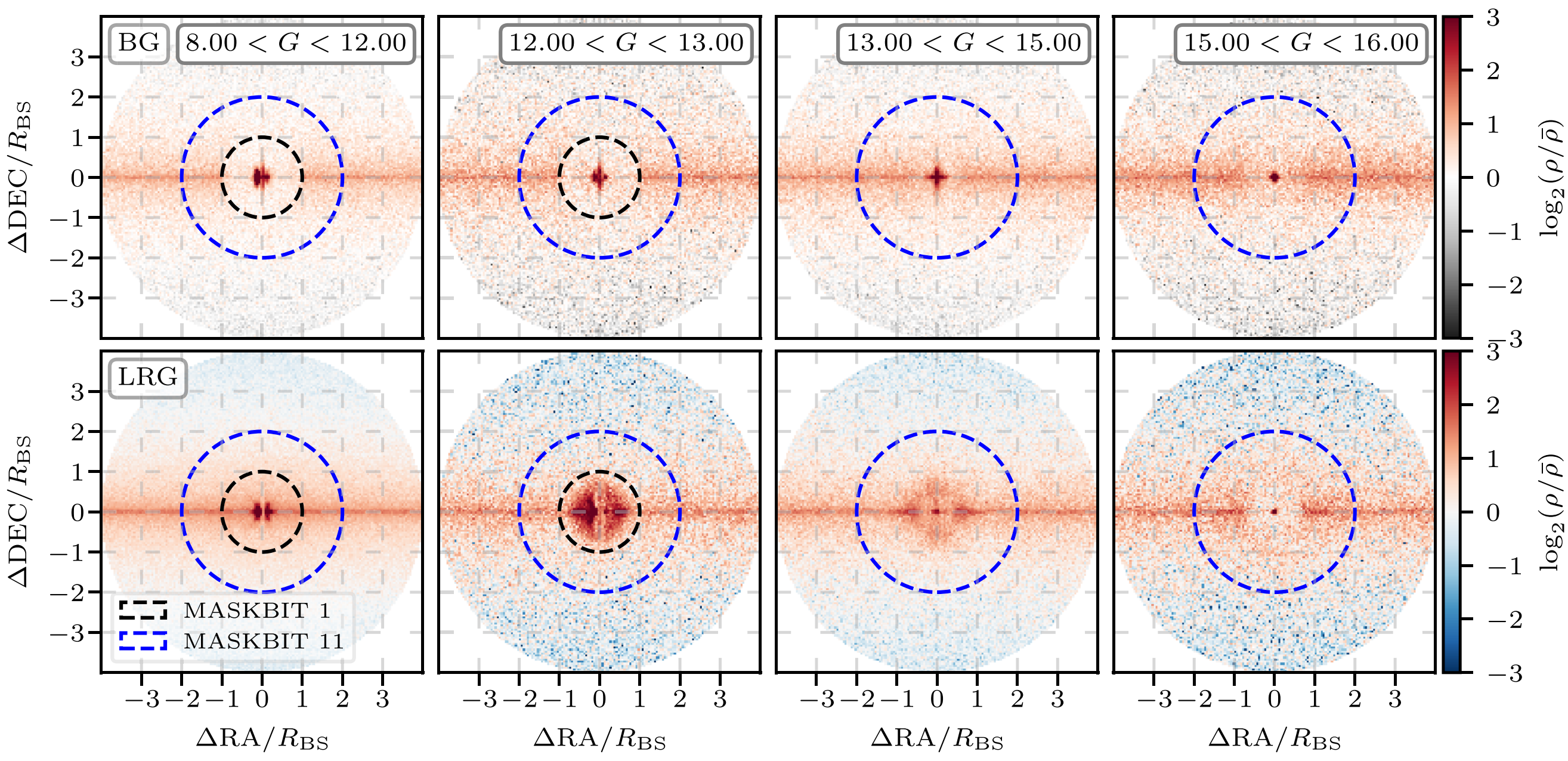

**Figure 3.** Stacked two-dimensional density maps of CRS BG (top row) and LRG (bottom row) targets around their nearest *Gaia* stars, split into four *Gaia* magnitude bins. ΔRA and ΔDec. show separations in RA and Dec. scaled by the bright-star mask radius $R_{\rm BS}$, respectively. The black circle marks the `MASKBIT 1` radius (available only for $G < 13$); the blue circle marks the `MASKBIT 11` radius, twice the `MASKBIT 1` radius, applied to all *Gaia* and Tycho 2 stars with $G < 16$. Colours indicate $\log_2(\rho/\bar{\rho})$, where $\rho$ is the per-pixel target density and $\bar{\rho}$ is the mean in the annulus $1.1 < R/R_{\rm BS} < 7$. The stacks show overdensities around stars with $G > 13$ in both tracers, supporting the use of `MASKBIT 11`. The thin near-horizontal features aligned with constant declination are bleed trails, not optical diffraction spikes.

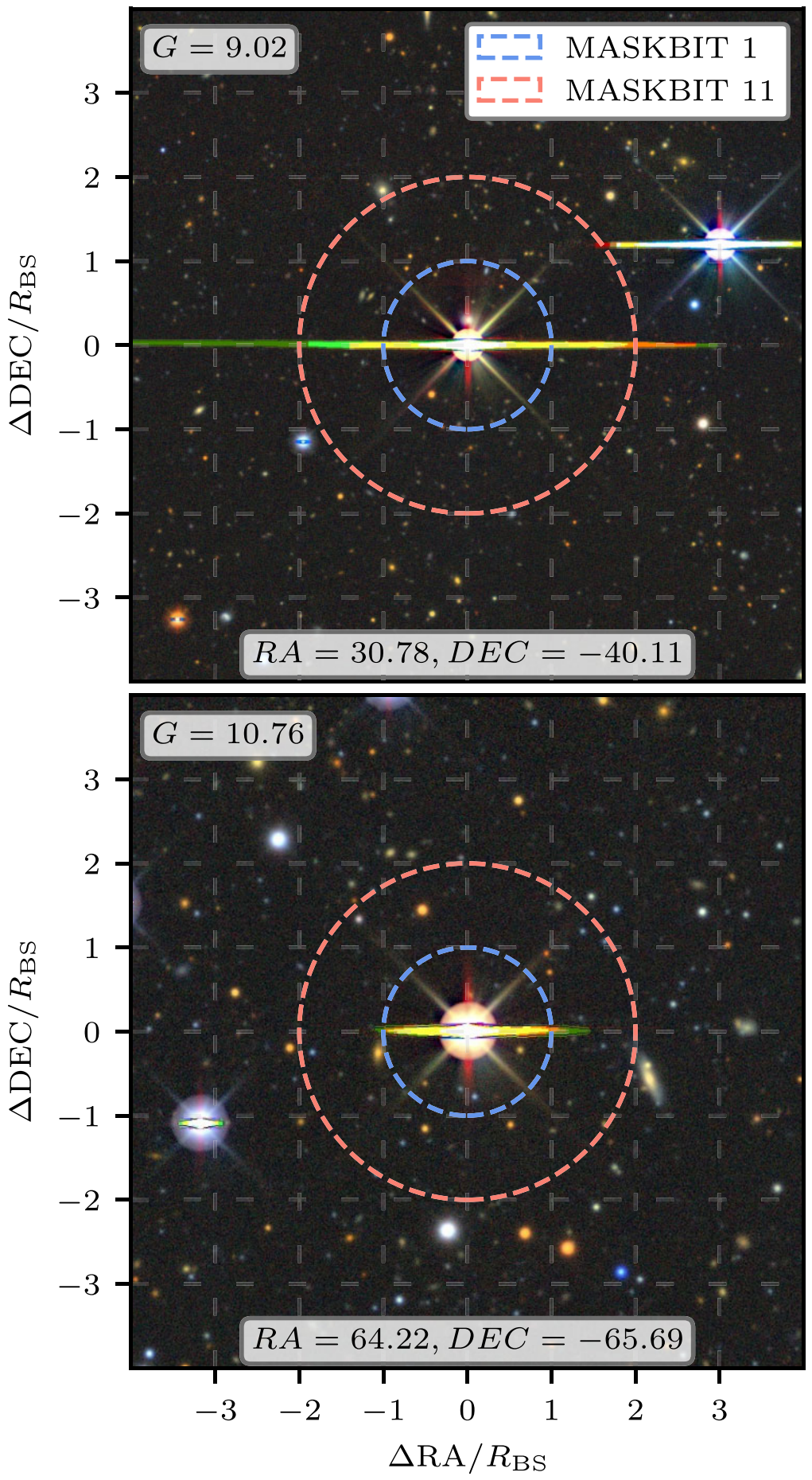

**Figure 4.** Examples of CCD saturation bleed trails (blooming) in the DESI Legacy Imaging Surveys, shown as cut-outs from the Legacy Surveys Sky Viewer around two randomly selected stars with $8 < G < 12$. The thin near-horizontal features at approximately constant declination are bleed trails, electronic artefacts distinct from the $\sim 45^\circ$ optical diffraction spikes, which are also visible. Axes show $\Delta\mathrm{RA}/R_{\mathrm{BS}}$ and $\Delta\mathrm{Dec}/R_{\mathrm{BS}}$, where $R_{\mathrm{BS}}$ is the bright–star mask radius. LS Sky Viewer Credit: Legacy Surveys/D. Lang (Perimeter Institute).

In this work, we do not apply photometric clustering weights to the CRS catalogues. Methods based on linear and random-forest regressions (e.g. E. Chaussidon et al. 2021) and their application to CRS are discussed by A. Verdier et al. (2025). A full exploration of such weights for CRS is left to future work; our focus here is to establish a masking configuration for which $w(\theta)$ is stable in the sense defined above, with $\Delta w(\theta)/\sigma(\theta)$ statistically consistent with zero on most scales.

[4] https://www.legacysurvey.org/viewer

## 4 LIMBER SCALING TEST FOR BRIGHT GALAXIES TARGET CATALOGUE

### 4.1 Limber's equation and the scaling test

Angular clustering is related to the spatial correlation function $\xi(r)$ through projection along the line of sight (LOS). D. N. Limber (1953) first derived the relation between $w(\theta)$ and $\xi(r)$ for a given redshift distribution of galaxies. The relativistic general form of the Limber's equation (initially derived by S. Phillipps et al. 1978) is given in Peebles (1980) as

$$w(\theta) = \frac{\int_0^\infty \int_0^\infty r_1^2 r_2^2 p(r_1)\, p(r_2)\, \xi(r_{12}, z)\, \mathrm{d}r_1\, \mathrm{d}r_2}{\left[\int_0^\infty r^2 p(r) \mathrm{d}r\right]^2}, \tag{10}$$

where $r_{12} = |\boldsymbol{r_1} - \boldsymbol{r_2}|$ and $p(r)$ is the selection function. In the special case of a power-law approximation, we can model the redshift-dependent spatial correlation function as

$$\xi(r, z) = \left(\frac{r_0}{r}\right)^{\gamma} (1+z)^{-(3+\varepsilon)}, \tag{11}$$

where $\gamma$ is the slope of the power law (identical to that in equation 2), and $\varepsilon$ is a parametrisation of clustering evolution. In this scheme, $\varepsilon = 0$ corresponds to the stable–clustering limit: bound pairs maintain (approximately) fixed physical separations, giving $\xi(r, z) \propto (1+z)^{-3}$ at fixed proper $r$ (see Peebles 1980, section 73). By contrast, taking $\varepsilon = \gamma - 3$ gives (approximately) constant clustering at fixed comoving separation; for $\gamma \simeq 1.7$ this is $\varepsilon \approx -1.3$ (S. J. Maddox, G. Efstathiou & W. J. Sutherland 1996). In Section 4.3, we discuss the effect of different $\varepsilon$ values on the Limber scaling.

Under the Limber approximation, which assumes small angles such that pairs of galaxies contributing to $w(\theta)$ lie at nearly the same redshift ($r_1 \approx r_2 >> r_{12}$), one can write $w(\theta)$ as

$$w(\theta) = \sqrt{\pi}\, \frac{\Gamma[(\gamma-1)/2]}{\Gamma(\gamma/2)}\, \frac{B}{\theta^{\gamma-1}} r_0^{\gamma}, \tag{12}$$

where

$$B = \frac{\int_0^\infty x^{5-\gamma} a^6 p^2(x) (1+z)^{(\gamma-3-\varepsilon)} F(x)^{-1}\, \mathrm{d}x}{\left[\int_0^\infty x^2 a^3 p(x) F(x)^{-1}\, \mathrm{d}x\right]^2}, \tag{13}$$

and $\Gamma(x) = \int_0^\infty t^{x-1} \mathrm{e}^{-t}\, \mathrm{d}t$ is the Gamma function. In equation (13), $x$ is the comoving distance at redshift $z$, $a$ is the scale factor, and $F$ comes from the metric, which for a flat universe is equal to unity (S. J. Maddox et al. 1996). The redshift distribution, $\mathrm{d}N/\mathrm{d}z$ is related to the selection function (G. Efstathiou et al. 1991) by

$$\left(\frac{\mathrm{d}N}{\mathrm{d}z}\right) \mathrm{d}z \propto x^3 a^3 \frac{p(x)}{F(x)} \left(\frac{\mathrm{d}x}{\mathrm{d}z}\right) \mathrm{d}z, \tag{14}$$

so we can rewrite equation (13) with the redshift distribution $dN/dz$ instead of the selection function as

$$B = \frac{\int_0^\infty x^{1-\gamma} (\mathrm{d}N/\mathrm{d}z)^2 F(x) (1+z)^{(\gamma-3-\varepsilon)} (\mathrm{d}z/\mathrm{d}x) \mathrm{d}z}{\left[\int_0^\infty (\mathrm{d}N/\mathrm{d}z) \mathrm{d}z\right]^2}. \tag{15}$$

E. J. Groth & P. J. E. Peebles (1977) introduced a powerful consistency check known as the Limber scaling test to verify whether

the angular clustering measurements across different magnitude-limited slices are consistent with a single underlying real-space $\xi(r)$. The idea is that if galaxies in various magnitude (or depth) slices share the same intrinsic clustering, then the observed $w(\theta)$ for each slice should correspond to the same $\xi(r)$ when properly scaled by the respective redshift distribution. In practice, one can use a fiducial real-space correlation (with parameters $r_0$, $\gamma$, and $\epsilon$) and the measured $N(z)$ of each slice to predict the expected $w(\theta)$ via Limber's equation. The scaling test involves comparing the measurements of $w(\theta)$ from different slices by shifting or scaling the curves according to these predictions. For example, a deeper (fainter) galaxy sample will have a lower $w(\theta)$ amplitude than a shallower (brighter) sample, due to the increased line-of-sight projection; the Limber equation quantitatively predicts this change in amplitude. By multiplying or dividing the $w(\theta)$ of one slice by the expected relative amplitude factors (and applying $\theta$-axis shifts for different effective depths), one can overlay the $w(\theta)$ curves from multiple slices. If the clustering is intrinsically the same, all slices should then collapse on to a single curve. Agreement within the uncertainties confirms that the observed differences in $w(\theta)$ are fully explained by the $N(z)$ variation, rather than by changes in the clustering or unaccounted systematic effects. This test, therefore, serves as a check that our angular clustering measurements truly represent the projected clustering of the three-dimensional galaxy distribution, and are not significantly biased by the sample selection or observational systematics (S. J. Maddox et al. 1996).

To align the $w(\theta)$ of different magnitude slices with a reference slice, we apply shifts in angle and amplitude, $\Delta \log_{10} \theta$ and $\Delta \log_{10} w$. We model the real-space correlation as a broken power law, $\xi(r) \propto r^{-\gamma_1}$ on small scales and $\xi(r) \propto r^{-\gamma_2}$ on large scales, which implies $w(\theta) \propto \theta^{1-\gamma}$ (with $\gamma = \gamma_1$ or $\gamma_2$ in the corresponding regimes). The resulting scaling factors are

$$\Delta \log_{10} \theta = \frac{\log_{10}\big[B_i(\gamma_1)/B_{\rm ref}(\gamma_1)\big] - \log_{10}\big[B_i(\gamma_2)/B_{\rm ref}(\gamma_2)\big]}{\gamma_2 - \gamma_1}, \quad (16)$$
$$\Delta \log_{10} w = (\gamma_1 - 1)\,\Delta \log_{10} \theta \;-\; \log_{10}\big[B_i(\gamma_1)/B_{\rm ref}(\gamma_1)\big],$$

where $\gamma_1$ and $\gamma_2$ are power-law fits slope for the reference slice, and $B$ is given by equation (15).

### 4.2 Redshift distribution and $\xi(r)$ models for BG catalogue

As discussed in Section 4.1, Limber scaling provides a robust test of consistency between angular clustering measurements in different magnitude slices. In this section, we apply the Limber scaling test to the CRS BG target selection.

To model the angular correlation function $w(\theta)$ using the Limber equation (equation 12), we require two key components: the redshift distribution $N(z)$ and the spatial correlation function $\xi(r)$. For the redshift distribution, we use the model introduced in C. M. Baugh & G. Efstathiou (1993, hereafter BE, equation 18):

$$\left(\frac{\mathrm{d}N}{\mathrm{d}z}\right) \mathrm{d}z \propto z^{\alpha} \exp\left[-\left(\frac{z}{z_c\,(m)}\right)^{\beta}\right] \mathrm{d}z, \quad (17)$$

where $m$ is the apparent magnitude. The upper panel of Fig. 5 shows the measured $N(z)$ for CRS-BG-like targets from DESI DR1 in different magnitude slices, along with the BE model fits. The fitted parameters are provided in Table 2. For comparison, the BE fits for the DESI BGS selection are also shown; the difference arises due to the colour cuts applied in the CRS BG selection to remove low-redshift ($z < 0.1$) galaxies, as detailed in VR25.

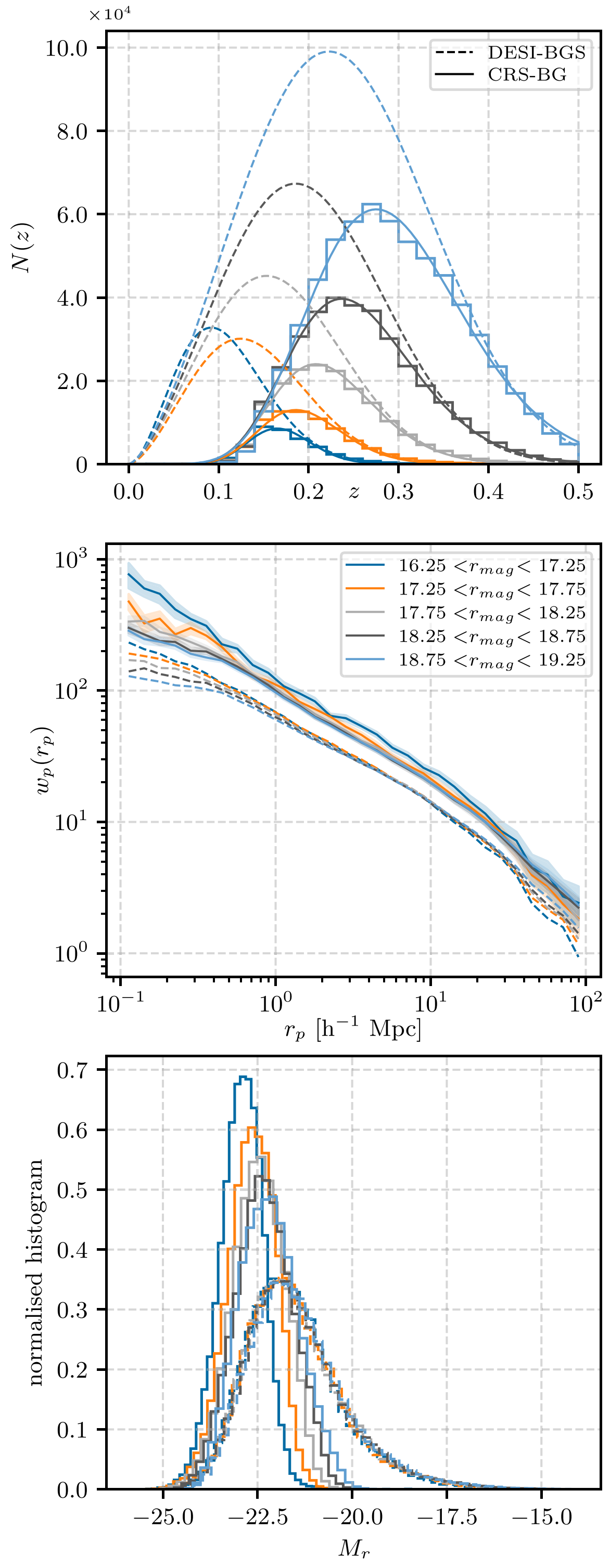

**Figure 5.** *Upper panel*: redshift distributions $N(z)$ of CRS BG-like targets from DESI DR1 in five $r$-band magnitude slices, with BE model fits (solid lines) and comparison to the DESI BGS sample (dashed line). *Middle panel*: projected correlation function $w_p(r_p)$ for the same slices, showing the scale-dependent clustering strength used to fit the spatial correlation parameters for Limber scaling. Error bars for $w_p(r_p)$ are calculated using 36 Jackknife regions. *Lower panel*: normalised histogram of absolute magnitude for CRS BG-like and DESI BGS. In the lower panel, absolute $r$-band magnitudes for DESI BGS are calculated using `Kcorrect v5` (M. R. Blanton & S. Roweis 2007).

**Table 2.** Best-fititng parameters of equation (18) (BE fit) for CRS BG and DESI BGS selections in DESI DR1.

| Selection | $r_{\mathrm{mag}}$ range | $\alpha$ | $\beta$ | $z_c$ |
|---|---|---|---|---|
| **CRS BG** | 16.25–17.25 | 13.511 | 1.633 | 0.045 |
| | 17.25–17.75 | 12.893 | 1.269 | 0.030 |
| | 17.75–18.25 | 11.976 | 1.126 | 0.025 |
| | 18.25–18.75 | 11.269 | 1.041 | 0.024 |
| | 18.75–19.25 | 8.367 | 1.259 | 0.061 |
| **DESI BGS** | 16.25–17.25 | 1.842 | 1.805 | 0.091 |
| | 17.25–17.75 | 1.603 | 2.199 | 0.143 |
| | 17.75–18.25 | 1.485 | 2.484 | 0.188 |
| | 18.25–18.75 | 1.483 | 2.616 | 0.231 |
| | 18.75–19.25 | 1.499 | 2.704 | 0.277 |

The spatial correlation function $\xi(r)$ is modelled using the projected correlation function $w_p(r_p)$, derived from the 2D redshift-space correlation function $\xi(r_p, \pi)$ via integration along the LOS (e.g. J. Loveday et al. 2018):

$$w_{\mathrm{p}}\left(r_p\right) = 2\int_0^{\pi_{\max}} \xi\left(r_p, \pi\right) \mathrm{d}\pi\,, \tag{19}$$

where $\pi$ is the LOS and $r_p$ is the projected separation. In this analysis, we calculated the integral using $\pi_{\max} = 50h^{-1}$Mpc. We verified that varying $\pi_{\max}$ between 50 and 100, $h^{-1}$Mpc changes the derived scaling factors by less than 0.1 per cent, with agreement to three significant figures for $r_p \leq 100$, $h^{-1}$Mpc.

The real-space correlation function $\xi(r)$ can be calculated using:

$$\xi_r(r) = -\frac{1}{\pi}\int_r^{\infty} w_{\mathrm{p}}\left(r_p\right)\left(r_p^2 - r^2\right)^{-1/2} \mathrm{d}r_p. \tag{20}$$

This approach avoids the effects of RSDs caused by peculiar velocities (A. L. Coil 2012), providing a cleaner measurement of real-space clustering.

To use equation (12), we require a power-law fit of $\xi(r)$, $\xi(r) = (r/r_0)^{-\gamma}$. Using equation (20), we fit a power law to $w_p(r_p)$ and estimate the correlation length $(r_0)$ and the slope of the power-law fit $(\gamma)$ using

$$\begin{aligned} w_{\mathrm{p}}\left(r_p\right) &= A r_p^{1-\gamma} \\ &= r_0^{\gamma}\sqrt{\pi}\left(\frac{\Gamma[(\gamma-1)/2]}{\Gamma(\gamma/2)}\right) r_p^{1-\gamma}, \end{aligned} \tag{21}$$

where $\Gamma$ is the gamma function (M. Davis & P. J. E. Peebles 1983).

The middle panel of Fig. 5 shows the projected correlation function $w_p(r_p)$ for CRS BG-like targets and DESI-BGS in DESI DR1, split by $r$-band magnitude. The corresponding clustering parameters from fits to equation (21) are listed in Table 3. The lower panel of Fig. 5 presents normalised absolute–magnitude histograms for CRS-BG and DESI-BGS. For CRS-BG, the mean absolute magnitude $\bar{M}_r$ is more negative (i.e. the sample is more luminous) and the dispersion $\sigma(M_r)$ is smaller than for DESI-BGS, owing to the additional colour selections described in VR25. As the upper panel shows, CRS-BG targets lie mainly within $0.1 \lesssim z \lesssim 0.5$; the selection therefore captures the more luminous subset of the DESI-BGS population. The clustering length $r_0$ is larger for CRS-BG than for DESI-BGS. The luminosity dependence of $w_p(r_p)$ reported by D. J. Farrow et al. (2015) based on the Galaxy And Mass Assembly Data Release II (GAMA DR II; J. Liske et al. 2015) and the Sloan Digital Sky Survey Data Release 7 (SDSS DR7; K. N. Abazajian et al. 2009) explains the difference between the CRS-BG and DESI-BGS $r_0$ values. Moreover, the colour cuts produce differing $M_r$ distributions across the CRS-BG magnitude slices (in contrast to the more similar DESI-BGS slices), which in turn accounts for the slice-to-slice variation in $r_0$ within CRS-BG.

**Table 3.** Clustering parameters and absolute magnitudes for CRS BG and DESI BGS selections in DESI DR1, split by $r$-band magnitude. A single power-law fit to $w_p(r_p)$ over $0.3 < r_p < 60\,h^{-1}$Mpc yields $r_0$, $\gamma$, and reduced $\chi^2_\nu$. Broken power-law fits over $0.3 < r_p < 30$ and $30 < r_p < 80\,h^{-1}$Mpc give $\gamma_1$ and $\gamma_2$.

| Selection | $r_{\mathrm{mag}}$ | $M_r$ range | $\bar{M}_r$ | $\sigma(M_r)$ | $N_{\mathrm{gal}}$ | $r_0$ [$h^{-1}$Mpc] | $\gamma$ | $\chi^2_\nu$ | $\gamma_1$ | $\gamma_2$ |
|---|---|---|---|---|---|---|---|---|---|---|
| **CRS BG** | 16.25–17.25 | [−25.43, −18.45] | −22.93 | 0.580 | 38 568 | $7.46 \pm 0.05$ | $1.75 \pm 0.01$ | 2.1 | $1.73 \pm 0.01$ | $2.38 \pm 0.48$ |
| | 17.25–17.75 | [−25.44, −16.81] | −22.66 | 0.640 | 75 350 | $6.88 \pm 0.04$ | $1.76 \pm 0.01$ | 2.6 | $1.74 \pm 0.01$ | $2.35 \pm 0.33$ |
| | 17.75–18.25 | [−25.27, −15.09] | −22.48 | 0.700 | 170 995 | $6.63 \pm 0.03$ | $1.75 \pm 0.01$ | 3.1 | $1.72 \pm 0.01$ | $2.16 \pm 0.25$ |
| | 18.25–18.75 | [−24.90, −15.58] | −22.33 | 0.750 | 343 187 | $6.50 \pm 0.03$ | $1.73 \pm 0.01$ | 2.7 | $1.72 \pm 0.01$ | $2.09 \pm 0.23$ |
| | 18.75–19.25 | [−24.41, −14.75] | −22.19 | 0.780 | 631 393 | $6.44 \pm 0.02$ | $1.71 \pm 0.01$ | 3.9 | $1.70 \pm 0.01$ | $2.02 \pm 0.19$ |
| **DESI BGS** | 16.25–17.25 | [−25.90, −16.05] | −21.57 | 1.240 | 201 452 | $5.19 \pm 0.05$ | $1.72 \pm 0.01$ | 1.3 | $1.70 \pm 0.01$ | $2.38 \pm 0.54$ |
| | 17.25–17.75 | [−25.44, −15.53] | −21.53 | 1.290 | 237 957 | $5.19 \pm 0.04$ | $1.70 \pm 0.01$ | 1.7 | $1.68 \pm 0.01$ | $2.23 \pm 0.44$ |
| | 17.75–18.25 | [−25.27, −15.02] | −21.49 | 1.320 | 427 375 | $5.13 \pm 0.03$ | $1.69 \pm 0.01$ | 3.0 | $1.66 \pm 0.01$ | $2.21 \pm 0.33$ |
| | 18.25–18.75 | [−24.90, −14.53] | −21.45 | 1.340 | 750 634 | $5.01 \pm 0.03$ | $1.68 \pm 0.01$ | 4.0 | $1.66 \pm 0.01$ | $2.21 \pm 0.26$ |
| | 18.75–19.25 | [−24.85, −14.02] | −21.38 | 1.340 | 1 283 318 | $4.93 \pm 0.02$ | $1.64 \pm 0.01$ | 6.3 | $1.62 \pm 0.01$ | $2.12 \pm 0.22$ |

**Table 4.** Scaling factors for CRS BG (DESI DR1) in $r$-band magnitude slices for $\varepsilon = 0$ and $\varepsilon = -1.3$. $\Delta \log w(\theta)$ and $\Delta \log \theta$ are the vertical and horizontal shifts applied relative to the reference slice. $A_w$ values are clustering amplitudes from a power-law fit (equation 2) over 0.15°–0.8°, shown before and after applying Limber scaling.

| $r_{\mathrm{mag}}$ range | Scaling factors | | | | $A_w$ (not scaled) $\times 10^{-2}$ | | $A_w$ (scaled) $\times 10^{-2}$ | |
|---|---|---|---|---|---|---|---|---|
| | $\varepsilon = 0$ | | $\varepsilon = -1.3$ | | | | | |
| | $\Delta \log \theta$ | $\Delta \log w(\theta)$ | $\Delta \log \theta$ | $\Delta \log w(\theta)$ | NGC | SGC | NGC | SGC |
| 16.25–17.25 | −0.122 | −0.512 | −0.124 | −0.478 | $7.75 \pm 0.83$ | $8.86 \pm 0.28$ | $1.85 \pm 0.23$ | $2.19 \pm 0.08$ |
| 17.25–17.75 | −0.080 | −0.321 | −0.082 | −0.297 | $4.33 \pm 0.36$ | $4.45 \pm 0.18$ | $1.74 \pm 0.16$ | $1.79 \pm 0.08$ |
| 17.75–18.25 | −0.042 | −0.161 | −0.043 | −0.148 | $2.64 \pm 0.22$ | $2.97 \pm 0.12$ | $1.64 \pm 0.15$ | $1.86 \pm 0.08$ |
| 18.25–18.75 | *reference slice* | | | | $1.63 \pm 0.09$ | $2.20 \pm 0.09$ | $1.63 \pm 0.09$ | $2.20 \pm 0.09$ |
| 18.75–19.25 | 0.045 | 0.152 | 0.046 | 0.139 | $1.22 \pm 0.04$ | $1.52 \pm 0.03$ | $1.94 \pm 0.02$ | $2.22 \pm 0.09$ |

### 4.3 Limber scaling test results

As shown in the middle panel of Fig. 5 and Table 3, the projected correlation function $w_p(r_p)$ departs from a single power law: some magnitude slices become shallower below $r_p \sim 1$ Mpc h$^{-1}$, and all slices fall below that power law on larger scales, which reflects the one-halo to two-halo transition. To account for this in the Limber projection, we adopt a broken power-law model for $\xi(r)$ with slopes $\gamma_1$ and $\gamma_2$ below and above a transition scale $r_b$, which improves the accuracy of the predicted Limber amplitude $B$ (equation 15) and hence the scaling shifts. The power-law form is essential for rescaling both the amplitude and the angular position of $w(\theta)$, since under the Limber approximation $w(\theta) \propto \theta^{1-\gamma}$. If $\xi(r)$ departs from a power law, the projected slope becomes slice-dependent and only amplitude scaling remains valid.

Using this model, we compute the horizontal and vertical scaling shifts $\Delta \log_{10} \theta$ and $\Delta \log_{10} w$ relative to a reference slice ($18.25 < r_{\mathrm{mag}} < 18.75$), as given by equations (16) and (17). We adopt $18.25 < r < 18.75$ as the reference because it is the most statistically robust, yielding the smallest jackknife uncertainties and the most stable $w(\theta)$ measurements across angular scales.

In this work, we use the magnitude-slice Limber scaling primarily as a check on the target catalogue systematics, rather than as a constraint on the redshift evolution of clustering or as evidence for redshift–invariant bias. For each magnitude slice, we predict the depth dependence of by inserting its measured value into equations (10) and (15), adopting a fiducial parametric evolution model characterised by the parameter . Because this prediction depends on assumptions about clustering and bias evolution, any overall offset in amplitude between the measured and predicted $w(\theta)$ can be absorbed into a different choice of evolution or bias model and cannot be uniquely interpreted as an imaging or selection systematic.

We therefore focus on the relative Limber scaling between magnitude slices. Applying the predicted shifts allows us to overlay the angular correlation functions from different slices and test whether the observed differences in $w(\theta)$ can be explained solely by projection effects, while being insensitive to any global amplitude offset. The corresponding scaling factors $\Delta \log_{10} \theta$ and $\Delta \log_{10} w$ are reported in Table 4 for both $\varepsilon = 0$ and $\varepsilon = -1.3$. Our measured Limber scaling offsets are essentially insensitive to the choice of $\varepsilon$, mirroring the conclusion of S. J. Maddox et al. (1996, table 5), who likewise found only weak dependence of the scaling factors on $\varepsilon$ at similar depths.

Fig. 6 shows $w(\theta)$ for the CRS BG targets in five $r$-band magnitude slices, before and after applying the Limber scaling with $\varepsilon = 0$. The test is performed separately for the North Galactic Cap (NGC; DECaLS) and South Galactic Cap (SGC; DECaLS + DES). Since these surveys differ in photometric depth, observing strategy, and systematics, comparing the two caps provides an additional uniformity test of the CRS BG target selection (bottom panels spanning the left and middle columns of Fig. 6); the right-hand panels provide a complementary within-cap check by showing the slice-to-reference residuals.

The middle column of Fig. 6 display the same measurements after applying the predicted horizontal and vertical shifts (equations 16 and 17) using the broken power-law $\xi(r)$ and the fitted $N(z)$ for each slice, with $18.25 < r < 18.75$ as the reference. Post-scaling, the $w(\theta)$ curves from all slices align closely over $0.05^\circ \lesssim \theta \lesssim 3^\circ$ in both caps, consistent with slice–to–slice differences being caused by projection through the respective $N(z)$.

This behaviour is expected if the measured angular signal is a projection of three-dimensional clustering through the slice-dependent selection, rather than being derived by spatially varying photometric systematics (S. J. Maddox et al. 1996). We therefore interpret the scaling test as a uniformity check, showing that the observed slice-to-slice differences are explained by $N(z)$; we do not assume or claim redshift-invariant intrinsic clustering.

As mentioned in Section 2.1.1, the CRS BG catalogue applies not only magnitude limits but also colour cuts to isolate galaxies within the desired redshift range and minimise stellar contamination. If these colour cuts introduced redshift- or magnitude-dependent selection effects (e.g. selecting different galaxy populations at different depths), the scaled $w(\theta)$ curves would diverge in amplitude or shape, even after accounting for differences in $N(z)$. Instead, the success of the Limber scaling test indicates that the colour selection has been applied consistently across slices and does not distort the underlying clustering signal. The galaxies selected in each magnitude bin appear to trace the same large-scale structure, supporting the reliability of the colour–magnitude selection strategy.

In summary, the Limber scaling test demonstrates internal consistency across magnitude slices and, via the cross-cap residuals $(w_{\mathrm{NGC}} - w_{\mathrm{SGC}})/\sigma_{\mathrm{cc}}$ in the bottom panels, agreement between NGC and SGC at the $\sim 1\sigma$ level after scaling, where $\sigma_{\mathrm{cc}}$ is the quadratic sum of the jackknife errors (equation 9). The right-hand panels provide an additional within-cap diagnostic by showing the slice-to-reference residuals. The observed differences in $w(\theta)$ are largely explained by variations in $N(z)$, and the scaled measurements exhibit coherence both within and between caps, supporting the robustness of the CRS BG target selection for cosmological clustering analyses.

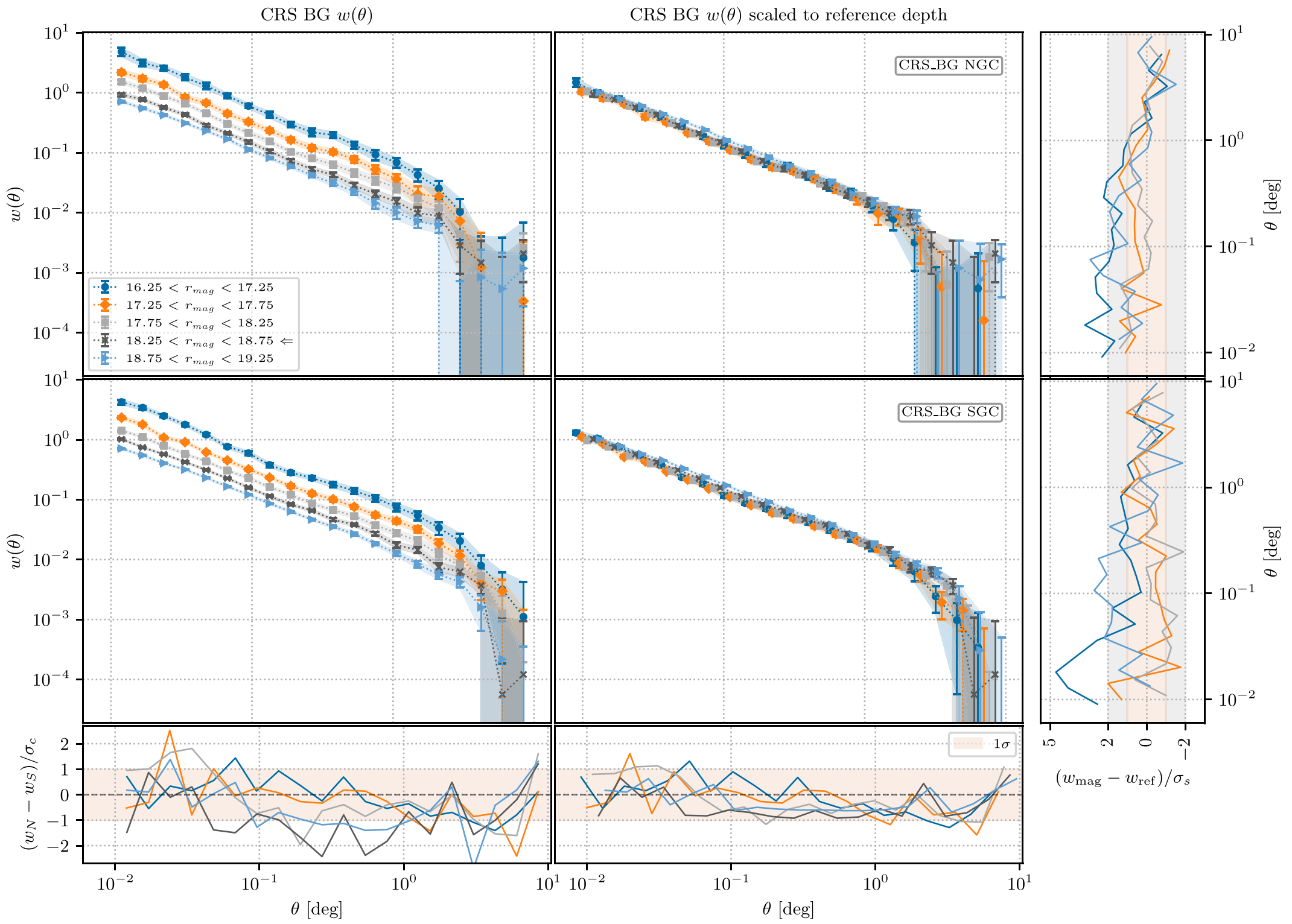

**Figure 6.** Angular correlation function $w(\theta)$ of the CRS BG targets in five $r$-band magnitude slices. *Left column:* measurements in the North Galactic Cap (NGC; DECaLS, top) and South Galactic Cap (SGC; DECaLS + DES, middle). *Middle column:* the same measurements after Limber scaling to the reference magnitude slice (as labelled). Error bars are estimated from 36 jackknife regions of comparable area across the CRS footprint. *Bottom row:* cross-cap residuals $(w_{\mathrm{NGC}} - w_{\mathrm{SGC}})/\sigma_{\mathrm{c}}$, where $\sigma_{\mathrm{c}}$ is the quadratic sum of the NGC and SGC jackknife errors. After scaling, curves from all magnitude slices overlay in both caps, and almost all points in the $\Delta w/\sigma$ panel lie within $\pm 1\sigma$. *Right column:* slice-to-reference residuals $(w_{\mathrm{mag}} - w_{\mathrm{ref}})/\sigma_{\mathrm{s}}$, where $\sigma_{\mathrm{s}}$ is the jackknife uncertainty on the difference (equation 9), shown as a diagnostic of remaining slice-to-slice differences after scaling.

As an alternative to the Limber-based projection test adopted here, one can assess the magnitude-slice behaviour of $w(\theta)$ using forward-modelled mock catalogues. In this approach, the identical target selection and imaging mask are applied to light-cone mocks, and are measured in the same magnitude slices and compared to the data after a common normalisation. This avoids any explicit assumption about clustering evolution in the projection (i.e. no $\epsilon$ is required), but it inherits the assumptions of the mock construction (e.g. halo occupation/abundance matching, colour–magnitude assignment, and the realism of imaging depth and artefact masks). A closely related strategy was used for the DESI Bright Galaxy Survey (e.g. P. Zarrouk et al. 2021), where the slice-to-slice scaling of the angular clustering was validated with mocks.

In this work, we do not present a mock-based cross-check for a pragmatic reason: at the time of analysis, suitably band-matched mocks for our selection were not yet available. Moreover, the P. Zarrouk et al. (2021) study predates DESI DR1 and therefore relied on pre-survey resources and limited spectroscopy. By contrast, we now have access to DESI DR1, which provides robust spectroscopic redshift information to constrain $N(z)$ and to benchmark any forward-model realisations. Incorporating a like-for-like mock validation, using the exact cuts and mask employed here and anchored to DESI DR1, is therefore a natural complementary test and will be pursued alongside our Limber exercise.

## 5 CROSS-CORRELATION WITH EXTERNAL SPECTROSCOPIC DATA AND $N(z)$

Clustering redshifts estimate the ensemble redshift distribution, $N(z)$, of a photometric sample by measuring its angular cross-correlation with a spectroscopic reference sample as a function of the reference redshift (J. A. Newman 2008; B. Ménard et al. 2014). The goal in this section is twofold. First, we validate the CRS BG redshift distribution by comparing the clustering-based estimate $P(z)$ to the directly observed $N(z)$ from DESI DR1 after applying the CRS BG selection. Second, we outline near-term applications of cross-correlations in the CRS overlap with deep imaging (e.g. LSST), where $N(z)$ calibration and measurements beyond spectroscopic limits are required. Operationally, we measure the angle and redshift-dependent cross-correlation between targets and a spectroscopic reference (DESI DR1), compress it to a function of redshift with an optimally weighted angular integral, and normalise by the reference autocorrelation to reduce nuisance dependences. We then compare the resulting $P(z)$ with the observed $N(z)$ in $r$-magnitude slices.

Unlike photometric redshifts, which estimate the redshift of individual galaxies using spectral energy distribution fits or machine learning (e.g. K. J. Duncan 2022; E. Tempel et al. 2025), clustering redshifts probe the *ensemble* redshift distribution of a population. This method is particularly robust to colour–redshift degeneracies, catastrophic photo-$z$ outliers, and photometric calibration errors, making it valuable for validating photometric selections and calibrating redshift distributions in cosmological analyses.

The technique involves binning the reference sample into narrow redshift intervals and computing the angular cross-correlation function between each slice and the full target sample. The resulting redshift-dependent clustering amplitude encodes the strength of overlap between the two populations at each redshift. In this way, clustering redshifts serves as a statistical probe of the redshift distribution, especially in regimes where direct spectroscopic measurements are observationally expensive or biased.

We apply clustering redshifts to the CRS BG sample and compare the resulting $N(z)$ shapes with those measured directly from DESI DR1 after applying the CRS-BG selection (Section 2.1). One application of Cluster-$z$ is to provide a reliable $N(z)$ in the absence of dense spectroscopy, in particular for the Limber scaling test discussed in Section 4. We also discuss further applications in Section 5.3.

### 5.1 Methodology of clustering redshift

The first step in the clustering redshift framework is to construct an appropriate reference sample with accurate spectroscopic redshifts and sufficient coverage in redshift and sky area to overlap with the target sample (C. B. Morrison et al. 2017). The target sample is typically drawn from a photometric catalogue and subdivided by observable quantities such as magnitude, colour, or photometric type.

The key observable is the *clustering amplitude*, $\bar{w}_t(z)$, which quantifies the integrated angular cross-correlation signal between the target and reference samples as a function of redshift. This is computed by integrating the angular cross-correlation function $w_t(\theta, z)$ over a specified angular range:

$$\bar{w}_t(z) = \int_{\theta_{\min}}^{\theta_{\max}} \mathrm{d}\theta \, W(\theta) \, w_t(\theta, z), \tag{22}$$

where $W(\theta)$ is a weight function; we adopt $W(\theta) \propto \theta^{-1}$, which is near-optimal for Poisson-dominated noise and a power-law correlation function (G. S. Karademir et al. 2021). The integration bounds, $\theta_{\min} = 0.005^{\circ}$ and $\theta_{\max} = 10^{\circ}$, match the angular range of our $w(\theta)$ measurements.

To translate the clustering amplitude into a redshift distribution, we assume that the observed signal is dominated by the overlap between the redshift distributions of the two samples. Under the further assumption that the galaxy bias and the matter correlation function vary slowly with redshift over the width of the redshift bins, the estimated redshift probability distribution $P(z)$ of the target sample (up to an overall normalisation) is given by

$$P_{m,z} \propto \frac{\bar{w}_{tr}}{\sqrt{\bar{w}_{rr} \Delta z}} \times \frac{1}{\bar{b}_t(z) \sqrt{\bar{w}_m(z)}}, \tag{23}$$

Here, $\bar{w}_{tr}$ is the integrated cross-correlation between the target and reference samples, while $\bar{w}_{rr}$ is the autocorrelation of the reference sample within redshift bins of width $\Delta z$. The terms $\bar{b}_t(z)$ and $\bar{w}_m(z)$ account for the redshift evolution of the galaxy bias of the target sample and the underlying matter clustering, respectively. In practical applications, these terms are often absorbed into a global scaling factor or marginalised over, under the assumption that they evolve slowly over the redshift bins of interest (e.g. B. Ménard et al. 2014; G. S. Karademir et al. 2021). While the method is statistically powerful and robust to many observational systematics, it is not without limitations. The clustering signal may be contaminated by foregrounds, spatial systematics, or masking effects, especially if these vary across the target sample. Additionally, the method recovers the *relative* redshift distribution and does not measure absolute number densities unless the target bias and selection function are independently calibrated. Nevertheless, clustering-based redshift estimation has proven to be a vital component of modern cosmological analyses, particularly in the context of weak lensing tomography, galaxy clustering, and photometric sample validation (e.g. M. Gatti et al. 2021; H. Hildebrandt et al. 2021).

### 5.2 $N(z)$ estimation for CRS-BG targets

Fig. 7 compares the clustering-derived redshift distributions (solid blue) with the DESI DR1 histograms after applying the CRS-BG selection (orange points) in six $r$-band magnitude bins between $16.25 < r < 19.25$. For each case, we overplot the BE fit to aid visual comparison. The BE fits are just applied on $P(z) \geq 0$. In this estimation, we used $\Delta z = 0.02$ and $\Delta m = 0.5$.

At fainter magnitudes ($r \gtrsim 17.5$), the shape closely tracks the spectroscopic after normalisation to unit area. This holds in both NGC and SGC. The faint-bin behaviour supports the use of clustering redshifts as an estimation of $N(z)$ for the Limber scaling test in the absence of observed spectroscopic redshift; however, in this work, we preferred using the observed $N(z)$ from DESI DR1. At the bright end ($r \lesssim 16.75$), the $P(z)$ curves show sharper features and larger bin-to-bin variations, mainly because that slice contains fewer galaxies, which lowers the cross-correlation S/N.

The clustering-redshift estimator returns a signed cross-correlation amplitude in each redshift bin, so in bins where the true signal is consistent with zero, it can fluctuate to negative values. This effect is most visible at the bright end and at very low redshift, where the CRS-BG selection removes most targets with $z < 0.1$ (see the top panel of Fig. 5), leaving only 0.2 per cent of DESI-BGS galaxies with $z$<0.1, which causes Jackknife to underestimate errors. The same thing happens for higher redshifts to have In these regimes, the cross-correlation signal is noise-dominated, and the reconstructed $P(z)$ can oscillate around zero, resulting in apparently negative probabilities. We interpret such negative excursions as statistical fluctuations around a vanishing signal rather than as physically meaningful negative densities, and they only occur where the expected contribution to the CRS-BG redshift distribution is already negligible.

When fitting an analytic template to the reconstructed $P(z)$ we therefore restrict the fit to the subset of points with $P(z) \geq 0$. For both the DESI DR1 histograms and the clustering-redshift estimates, we adopt the BE fit form as a smooth positive single-peaked model for the redshift distribution of a magnitude-limited galaxy sample. This choice enforces positivity, suppresses bin-to-bin noise, and allows a homogeneous comparison between the spectroscopic and clustering-derived $N(z)$ in terms of a small set of parameters. The BE fits are used here as a convenient summarising model and for visual comparison only: wherever spectroscopic redshifts are available, the Limber-scaling analysis in this work employs the observed DESI $N(z)$ rather than the fitted cluster-$z$ template, so our main results do not depend on the detailed choice of functional form.

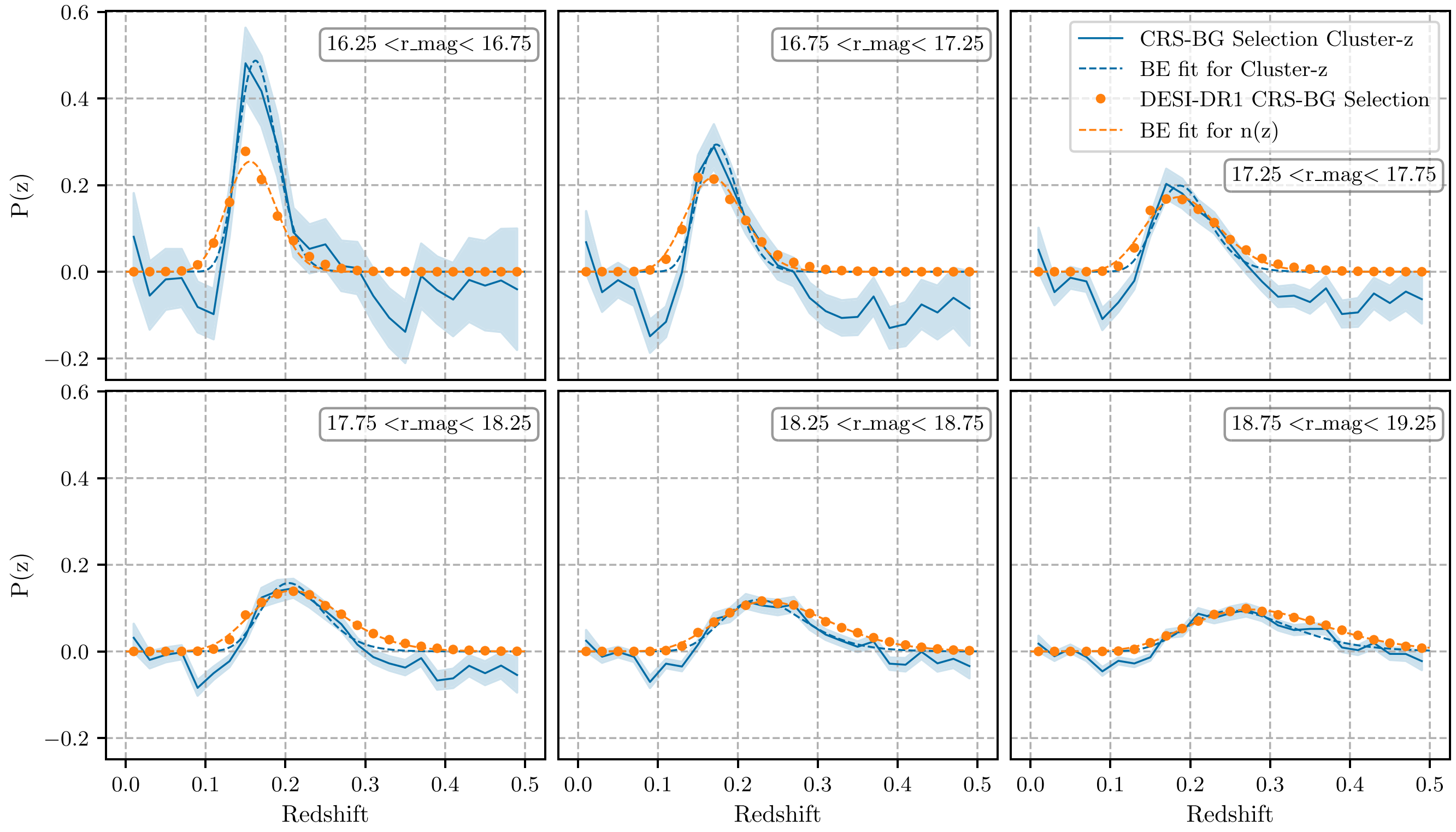

**Figure 7.** Comparison of normalised redshift distributions for CRS BG targets in six $r$-band magnitude bins from $16.25 < r < 19.25$. The orange dots represent the observed redshift histograms from DESI DR1 with CRS-BG selection; the dashed orange lines show Baugh & Efstathiou (BE) fits to these points. The solid blue lines show redshift distributions estimated using the clustering redshift (cluster-$z$) technique, and the dashed blue lines are BE fits to the cluster-$z$ results. At fainter magnitudes ($r \gtrsim 17.5$), the cluster-$z$ signal closely tracks the spectroscopic $N(z)$, validating the CRS selection. At brighter magnitudes, deviations reflect increased noise, local clustering, and possible contamination or cross-correlation limitations.

Overall, the results confirm that clustering redshifts are robust and accurate for the CRS BG sample at intermediate and faint magnitudes. They provide an important validation of the target selection and redshift distribution modelling in this regime. At brighter magnitudes, caution is warranted, and future improvements may involve refined cross-correlation strategies or auxiliary validation using deeper background samples.

## 5.3 Future applications enabled by survey overlaps

The extensive overlap of 4MOST-CRS with the LSST and *Euclid* (Ž. Ivezić et al. 2019; Euclid Collaboration 2025, see Fig. 1) enables cross-correlation analyses that are complementary to photometric approaches. In this context, clustering-based methods use position cross-correlations between CRS spectroscopic slices and wide photometric samples to infer, validate, or refine the redshift distributions required for weak-lensing and clustering tomography. The same framework also extends measurements beyond spectroscopic limits by supplying ensemble redshift information for faint populations.

*Calibration of tomographic $N(z)$ for weak lensing and clustering:* The CRS×LSST and CRS×Euclid footprints allow the calibration of photometric tomographic source-bin redshift distributions by cross-correlating each bin with CRS spectroscopic slices. This external calibration path is independent of photo-$z$ training and is sensitive to actual sky overlap, which makes it well suited to southern surveys. Recent studies quantify that clustering-redshift calibration can reach, and in some cases exceed, the accuracy targets set for Stage IV surveys when systematic effects such as magnification and redshift-dependent bias are modelled or controlled (e.g. M. Gatti et al. 2021).

*Extending measurements beyond spectroscopic limits:* Within the CRS×LSST area, clustering-based redshifts can provide $N(z)$ estimates for samples fainter than the magnitude limit of the reference catalogue, but not beyond the redshift limit of the reference sample (CRS redshift limit), enabling luminosity and stellar mass function measurements that leverage deep LSST photometry with CRS as the spectroscopic backbone. Practical designs combining magnitude-binned $P(z)$ with forward models of selection and completeness have already demonstrated feasibility for pushing to much lower luminosities than direct spectroscopy alone (see G. S. Karademir et al. 2021, 2023).

*Photometric-redshift calibration for* Euclid *and LSST:* Cross-correlation methods can also calibrate photo-$z$ directly by constraining both the mean redshift and the shape of $n(z)$ for photometric samples, either as priors on photo-$z$ hyper-parameters or within joint likelihoods that combine clustering and photometry. Forecasts and simulation-based studies for *Euclid* show that cross-correlation calibration meets the required precision on bin means provided key systematics are accounted for, and the same approach applies to LSST over the common footprint with CRS (see K. Naidoo et al. 2023; W. D. Doumerg et al. 2025). Together with Fig. 1, these overlaps motivate a unified CRS-based calibration strategy for both surveys.

**Table 5.** Best-fititng parameters of equation (2) for 4MOST LRGs in the DES and DECaLS region for different photometric redshift slices.

| $z$ bins | DECaLS $A_\omega$ | DECaLS $\gamma$ | DECaLS $\chi^2_\nu$ | DES $A_\omega$ | DES $\gamma$ | DES $\chi^2_\nu$ |
|---|---|---|---|---|---|---|
| $0.4 < z < 0.5$ | 0.029 | 1.837 | 1.202 | 0.031 | 1.821 | 0.566 |
| $0.5 < z < 0.6$ | 0.031 | 1.851 | 0.790 | 0.031 | 1.856 | 0.336 |
| $0.6 < z < 0.7$ | 0.026 | 1.869 | 0.809 | 0.028 | 1.871 | 1.726 |
| $0.7 < z < 0.8$ | 0.021 | 1.974 | 0.953 | 0.021 | 1.997 | 1.182 |
| $0.8 < z < 0.9$ | 0.016 | 1.998 | 1.672 | 0.018 | 2.023 | 2.051 |
| $0.9 < z < 1.0$ | 0.015 | 1.930 | 0.462 | 0.015 | 2.057 | 0.952 |

The same framework applies to LRG targets. The bias $b_t(z)$ evolves appreciably with redshift. Consequently, the $1/(\bar{b}_t(z)\sqrt{\bar{w}_m(z)})$ factor in equation (23) cannot be treated as constant. We therefore require a realistic model for $b_t(z)$; a full LRG clustering-redshift analysis is left to future work.

## 6 ANGULAR CLUSTERING OF LRG TARGET SAMPLE

To test CRS-LRG target selection, we perform fits using a power-law model described in equation (2) to the angular correlation function in different redshift bins. The redshift distribution of LRGs can be found in our companion paper (VR25, fig. 7). The fitting results are reported in Table 5 and Fig. 8.

The angular clustering of LRGs is studied in six redshift bins of width 0.1 in the range $0.4 < z < 1$. The redshift of the targets is obtained using a Random Forest algorithm described in R. Zhou et al. (2021). As the photometry is deeper in the DES footprint compared to DECaLS, the evaluation of the angular 2CPF is split accordingly. The angular 2PCF is calculated between $0.01^\circ$ and $10^\circ$ using 41 logarithmic bins. The error bars were evaluated using the Jackknife resampling method with $N = 36$ subregions. The power-law model is compared to the angular clustering between scales $0.1^\circ < \theta < 0.8^\circ$ using the following $\chi^2$ definition:

$$\chi^2 = \sum_{\theta_i} \frac{[\omega_{\rm data}(\theta_i) - \omega_{\rm PL}(\theta_i)]^2}{\sigma^2_{{\rm jk},\theta_i}}, \tag{24}$$

where $\omega_{\rm data}(\theta)$ and $\sigma^2_{\rm jk}$ are the angular 2PCF measurements and the corresponding Jackknife errors. $\omega_{\rm PL}(\theta)$ the prediction from the power-law model. The minimisation is performed using the `scipy curve_fit` method based on the least-squares algorithm. The fitting results for each region are reported in Table 5 and Fig. 8. The power-law model gives good fits to the data, as reported by the $\chi^2$ values in Table 5. Only the redshift bin $0.8 < z < 0.9$ reports a high $\chi^2$ value for both photometric regions. This can indicate potential small contaminations in the sample for this particular redshift bin. The power-law index $\gamma$ increases with redshift while the amplitude $A_\omega$ tends to decrease. The values of $\gamma$ are of the same order (slightly lower) than previous LRGs studies (U. Sawangwit et al. 2011) that reported a value of $\gamma \sim 2$ for different LRG samples. The small difference is most likely due to the difference in target selection. To conclude, the angular clustering of the CRS-LRG sample follows Limber's approximation for a power-law model at intermediate scales, indicating small contamination in the selected LRG sample for both photometric regions. In addition, the bottom panel of Fig. 8 shows, for each redshift bin, the error-normalised residual between the DECaLS and DES angular correlation functions, defined as

$$\frac{\omega_{\rm DECaLS} - \omega_{\rm DES}}{\sqrt{\sigma^2_{\rm DECaLS} + \sigma^2_{\rm DES}}},$$

assuming independent uncertainties. The difference is higher at low angular separation $\theta < 1^\circ$ between two regions up to $3\sigma$ differences compared to the jackknife uncertainties, while at large separation angles the differences lie within $1\sigma$. These reflect the impact of the quality of the photometry between these two regions.

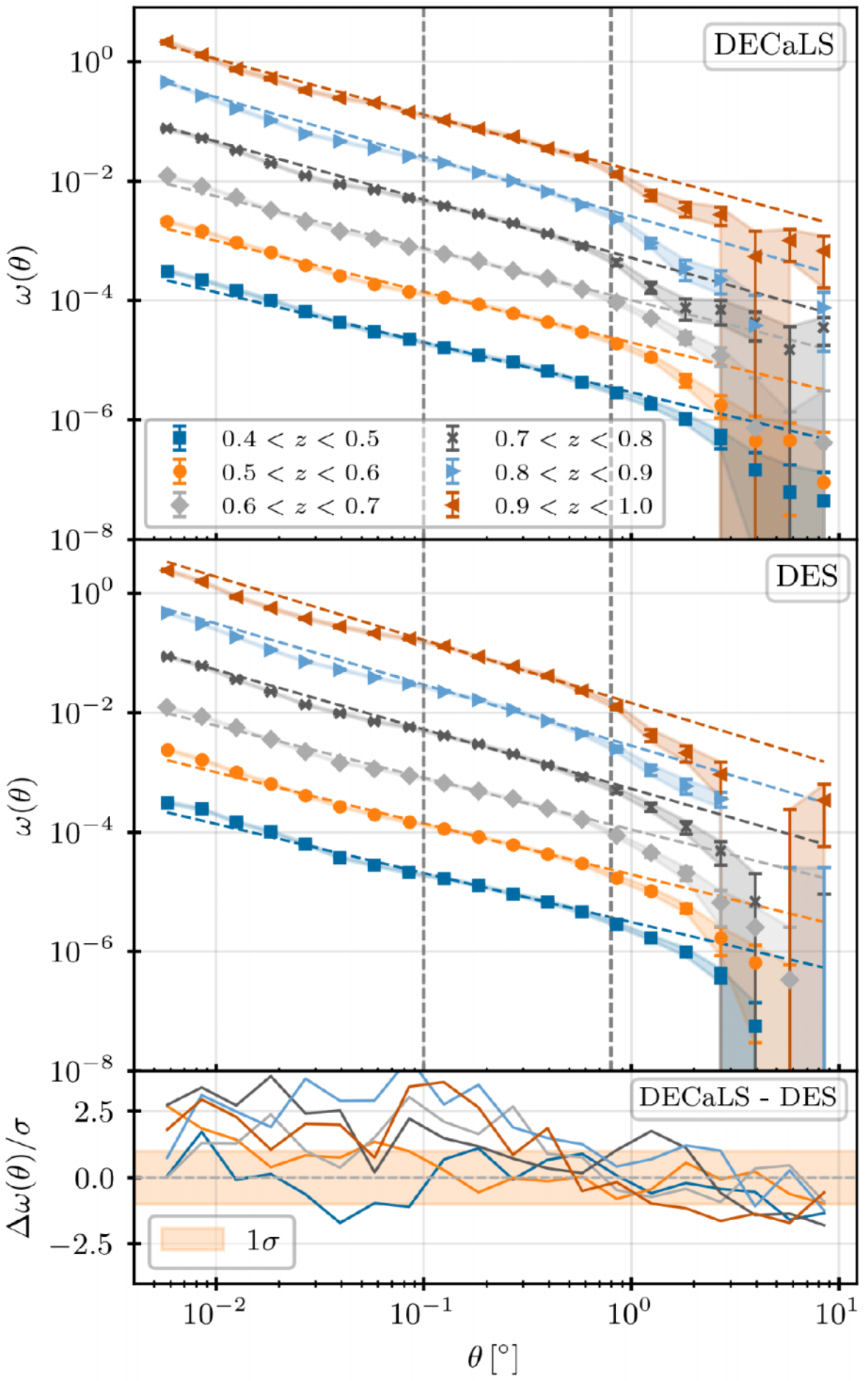

**Figure 8.** *Top panel:* the angular correlation function of the CRS-LRGs sample in the DECaLS footprint is shown as points with error bars. For clarity, each line (except for the highest redshift bin) has been scaled down to avoid overlap between the curves. Each colour corresponds to a photometric redshift bin, with a width of 0.1, ranging from 0.4 to 1.0. The dashed line shows the best fit of the power-law model. The error bars were obtained from 36 jackknife subregions. The vertical lines show the fitting range used for the fit. *Middle panel:* same as the upper panel for the DES photometric region. *Bottom panel:* a comparison of the DECaLS and DES photometric regions. The lines show the difference between the angular correlation $\omega_\theta$ in the DECaLS and DES region, normalised by the quadratic sum of the jackknife errors. Each colour corresponds to a different redshift bin and uses the same colour code as the first two panels. The shaded orange region represents the $\pm 1\sigma$ band.

**Table 6.** Uniform prior ranges are used for the HOD minimisation.

| Parameters | Prior range |
|---|---|
| $\log_{10}(M_0)$ | [12.00, 14.00] |
| $\log_{10}(M_1)$ | [13.00, 14.50] |
| $\alpha$ | [0.70, 1.40] |
| $\log_{10}(M_{\rm cen})$ | [12.00, 14.00] |
| $\sigma_M$ | [0.05, 1.00] |

Across $0.4 < z < 1.0$, the CRS-LRG angular clustering is well described by a single power-law on $0.1^\circ < \theta < 0.8^\circ$ in both DES and DECaLS. The bin $0.8 < z < 0.9$ shows elevated $\chi^2$, hinting at minor contamination or residual imaging systematics, but the impact is confined mainly to small angular scales. For analyses sensitive to small scales, adopting conservative cuts (e.g. $\theta > 0.2^\circ$) or light systematics weighting is prudent; for large-scale applications, the target selection appears robust, with cross–footprint agreement at the $\lesssim 1\sigma$ level.

## 7 HOD FITTING OF PROJECTED CORRELATION FUNCTION OF LRG TARGETS

This section aims to give a description of the galaxy-halo connection for the CRS-LRG sample using the Halo Occupation Distribution (HOD) model (Z. Zheng et al. 2005). The HOD model is an empirical model that populates galaxies in dark matter haloes from $N$-body simulations. Studying this connection allows us to get a description of the galaxy sample and its clustering properties, such as the host halo population and the large-scale galaxy bias, that can be used to perform forecasts of the BAO/RSD constraints (see R. H. Wechsler & J. L. Tinker 2018 for a review). The LRG sample is divided into six redshift bins of width $\Delta z = 0.1$ between 0.4 and 1, similarly to Section 6. The photometric redshifts are predicted using a Random Forest algorithm described in R. Zhou et al. (2021). As the quality of the photometry depends on the different surveys/regions of the legacy surveys, with deeper photometry in the DES region (see VR25), we separate and compare the clustering measurements between these two regions and the full CRS-LRG sample. The projected clustering $w_p(r_p)$ (defined in equation 20) is evaluated in the range $0.01\ {\rm Mpc}\,h^{-1} < r_p < 30\ {\rm Mpc}\,h^{-1}$ using 20 logarithmic bins and 300 linear $\pi$ bins between $[-150,150]\ {\rm Mpc}\,h^{-1}$. The error bars were evaluated using the Jackknife resampling method (C.-F. J. Wu 1986) with $N = 36$ subregions.

### 7.1 HOD modelling

To model the late-time matter field, we use $N$-body simulations from the AbacusSummit suite, which use the CompaSO halo finder B. Hadzhiyska et al. (2021) to obtain the halo catalogues. We use the `highbase` simulation box with the baseline cosmology Planck 2018 ΛCDM (Planck Collaboration VI 2020): $\omega_{\rm cdm} = 0.12$, $\omega_b = 0.02237$, $h = 0.6736$, $\sigma_8 = 0.807952$, and $n_s = 0.9649$. The size of the cubic simulation box is $1\ {\rm Gpc}\cdot h^{-1}$ with a mass resolution of $2 \cdot 10^9 M_\odot/h$. The simulation boxes are taken at redshifts 0.45, 0.575, 0.65, 0.725, 0.875, and 0.95 corresponding to each redshift bin from 0.4 to 1. We use the standard HOD model (Z. Zheng et al. 2005) to fit the projected clustering of CRS-LRGs. The HOD is divided into two functional forms that describe the mean occupation number of galaxies according to the host halo mass $M_h$. One for the central galaxy occupation $\langle N_{\rm cen}(M_h)\rangle$ and one for the satellite occupation $\langle N_{\rm sat}(M_h)\rangle$. The central galaxy probability is given by a step-like function:

$$\langle N_{\rm cen}(M_h)\rangle = \frac{1}{2}\left[1 + {\rm erf}\left(\frac{\log_{10} M_h - \log_{10} M_{\rm cen}}{\sigma_M}\right)\right], \qquad (25)$$

$M_{\rm cen}$ determine the minimum halo mass and $\sigma_M$ the steepness of the step function. The satellite's occupation is described by a power law:

$$\langle N_{\rm sat}(M_h)\rangle = \left(\frac{M_h - M_0}{M_1}\right)^\alpha \qquad (26)$$

$M_0$ describe the minimum halo mass that can host satellite galaxies, characterises the typical halo mass from which you expect to host satellite galaxies, and $\alpha$ is the power-law index. The mean numbers of galaxies are then turned into a deterministic number for each halo using a Bernoulli distribution for central galaxies and a Poisson distribution for satellite galaxies. We allow satellite galaxies to populate haloes with no central galaxies. The positions of satellite galaxies within their host haloes are assumed to follow the Navarro–Frenk–White (NFW) profile (J. F. Navarro, C. S. Frenk & S. D. M. White 1997). The concentration parameter $c = R_h/R_s$ is computed using the Abacus simulation outputs, with $r_{98}$ taken to be the halo radius $R_h$ and $r_{25}$ the scale radius $R_s$, as described in A. Rocher et al. (2023a). $r_{98}$ and $r_{25}$ are the radii enclosing 98 per cent and 25 per cent of the halo particles. We do not apply any constraints to the mock galaxy density, but rather rescale the mean central and satellite occupation numbers, $\langle N_{\rm cen}(M_h)\rangle$ and $\langle N_{\rm sat}(M_h)\rangle$, by a factor $A \le 1$, which changes their amplitude. This means that the central occupation of the LRG may not reach 1 at high halo mass, with $A$ accounting for the incompleteness of the selected sample.

The generation of mock galaxy catalogues using the HOD model is performed using the python package `HODDIES` [5] (A. Rocher et al. 2023b). We use `pycorr`, a Python wrapper of the Corrfunc software package M. Sinha & L. Garrison (2019a), to measure the projected clustering $w_p(r_p)$ of the mock galaxies.

The inaccuracy in the photo-$z$ estimate will induce a bias in the clustering measurements. The photo errors will effectively randomise the galaxy distribution along the LOS, i.e. pairs of galaxies can be lost due to one of the galaxies being outside of the redshift bin, resulting in a lower amplitude than the true clustering signal measured with spectroscopic redshifts. To account for the photo-$z$ errors in the model, we perturb the observed position of the mock galaxies along the LOS by adding a smearing effect to their velocities. This effect is drawn from a Gaussian distribution of width taken from the mean of the photo-$z$ errors of the LRGs in the corresponding redshift bin, rescaled by a scaling factor $S_z$.

[5] https://hoddies.readthedocs.io

This scaling factor is introduced to account for uncertainties in the photo-$z$ error estimation as in R. Zhou et al. (2021), but we fix this value instead of fitting it. Based on the results from R. Zhou et al. (2021), we set this rescaling factor to 0.7 for the first two redshift bins and 0.6 for the higher redshift bins. The values of the smearing effect and $S_z$ are reported in Table 7.

### 7.2 HOD minimisation procedure

The fits are performed using the HOD formulation from equation (25) and (26) with five free parameters: $\log M_{\rm cen}$, $M_0$, $M_1$ $\sigma_M$, and $\alpha$. We employ `stochopy`,[6] a Python package for stochastic minimisation using Covariance Matrix Adaptation – Evolution Strategy. While the minimisation algorithm can provide fast estimation of the best-fitting parameters, it does not provide reliable error estimates. We do not perform a full Bayesian analysis using MCMC, and only provide the best-fitting from the minimisation procedure. Therefore, we do not quote error bars, and the reported results are mainly to perform a qualitative check of the sample and comparison with previous studies. We used uniform priors to perform the minimisation reported in the table, and the initialisation point is taken as the middle of the prior range. The fitting range is chosen to be between $r_p < 10$ Mpc/$h$ to avoid potential imaging systematic effects on larger scales. We use the $\chi^2$ function to minimise during the fitting procedure, defined as

$$\chi^2 = [w_p^{\rm mock} - w_p^{\rm data}] \cdot {\rm Cov}_{\rm JK}^{-1} \cdot [w_p^{\rm mock} - w_p^{\rm data}]^T, \quad (27)$$

where $w_p^{\rm data}$ and $w_p^{\rm mock}$ are the projected clustering measurements from the data and the HOD mock. ${\rm Cov}^{-1}$ is the inverse of the jackknife covariance matrix. We neglect the stochastic behaviour of the HOD model, since these errors are subdominant compared to the JK errors. We then perform minimisation by fixing the random seed.

### 7.3 HOD results

Fig. 9 displays the projected correlation function of the CRS-LRG sample in the six redshift bins with their corresponding best-fititng HOD results in three different cases: for the full CRS-LRG sample (labelled as 'ALL') and each of the photometric regions, DES/DECaLS. The dotted lines represent the clean clustering cases where no smearing effect is applied to mimic the effect of photo-$z$ errors. The corresponding best-fititng parameters are reported in Table 7. We take advantage of the DESI DR1 public data to create a spectroscopic sample using the CRS-LRG photometric selection. To avoid regions with low fibre completeness, we use an extra cut to the number of overlapping tiles `NTILE` $> 3$. Fig. 11 presents the comparison of the projected clustering measurements $w_p$ of the CRS-LRG spectroscopic sample to those of the DESI LRGs. As expected from target selection, the largest differences are at high redshifts, where the CRS-LRG sample selects brighter objects compared to the DESI-LRG sample (see VR25), resulting in an increase in the amplitude of the clustering signal.

The minimisation results shown in Fig. 9 are in decent agreement with CRS-LRG clustering data for all redshift bins, as shown by the residuals in Fig. 9, where the fits are mostly within compared to the data. However, the reduced $\chi^2$ value reported in the minimisation results (Table 7) is large. We associate this to the JK estimate of the covariance, which can be noisy in cross-correlation terms. We report in parentheses the corresponding $\chi^2_{\rm red}$ only using the JK variance. When the smearing effect is removed in the mocks, the clustering is closer to the CRS-LRG spec-$z$ sample across the redshift range considered. The values chosen for the rescaling factor $S_z$ to correct photo-$z$ errors seem to be valid with what one would expect with the spectroscopic sample. We note that the redshift bin $0.7 < z < 0.8$ seems to have a more precise photo-$z$ since the change in clustering amplitude is small between the spec-$z$ and photo-$z$ LRG samples.

The HOD results are difficult to interpret, as we do not derive errors from the minimisation procedure. However, we can draw qualitative trends and comparisons with other studies. From Table 7, we first notice that the satellite fraction of the samples remains stable across redshift around $\sim 10$ per cent, consistent with previous LRG HOD analysis (Z. Zhai et al. 2017; R. Zhou et al. 2021; S. Yuan et al. 2022, 2023). The value of the power law index $\alpha$ and $\log_{10}(M_1)$ remains around one and 14 $\log_{10}(M_\odot/h)$ for all redshifts, similarly to the LRG sample from the DESI 1 per cent survey results (S. Yuan et al. 2023). The minimum halo mass that can host a satellite is lower than or equal to $\log_{10}(M_{\rm cen})$ in almost every case, but few of these haloes will host LRG satellites given the value of $M_1 \sim 10^{14} M_\odot/h$. The value of $\log_{10}(M_{\rm cen})$ and $\sigma_M$ can differ by a few decimals between the samples, but the degeneracy between these two parameters leads to similar mean halo mass $\log_{10}(\overline{M_h})$. There is no clear trend in the mean halo mass of the CRS-LRG sample across redshift, but we report similar behaviour to DESI-LRG at high redshift, namely the mean halo mass of the sample tends to be lower (S. Yuan et al. 2023). This can result from the target selection that generates a drop in the redshift distribution at redshift $z \sim 1$, and might result in a physically different sample than the lower redshift LRG sample.

Finally, we derive the predicted linear bias factor of the CRS-LRG sample. To do so, we produce 50 mocks with HOD parameters randomly selected around the best-fititng values. As the minimisation procedure does not provide confident errors, we allow variation of the HOD best parameters in a range of $\pm 0.5$ for $\log_{10}(M_0)$ and $\pm 0.1$ for the other 4 parameters. These ranges are chosen to represent a qualitative and conservative estimate of the potential errors from a full Bayesian procedure. We then compute the real-space 2PCF monopole from these mocks and compare it to the predicted linear matter 2PCF monopole from linear theory (at the same cosmology), which is related by the squared value of the linear bias factor of the galaxy sample:

$$\xi_{gg}(r) = b_{\rm lin}^2 \xi_{mm}^{\rm lin}(r) \quad (28)$$

The linear matter 2PCF is derived using the Python package `cosmoprimo`[7] based on the Boltzmann code CLASS (J. Lesgourgues 2011). We evaluate equation (28) for scales between 40 and 80 Mpc $h^{-1}$ and fit the value of b for each of the 50 mocks. We then report the mean and the dispersion over the mocks of the measured linear bias in Fig. 10. There are no significant deviations in the inferred bias values from the two photometric regions across the redshift bins. These values are compared to the redshift evolution of the inverse of the linear growth factor as $b^{\rm lin} \propto 1.5/D(z)$. The bias reported in this study evolves consistently with the growth factor, as observed in previous LRG photometric studies (see R. Zhou et al. 2021). However, the lowest redshift bins appear to exhibit a lower bias value, which is unexpected given that the CRS-LRG selection is comparable to the DESI selection at these redshifts, as illustrated in Fig. 11, where the projected clustering amplitude of both CRS-LRG and DESI-LRG samples has the same amplitude. This trend is more likely due to the correction of the photo-$z$ error estimate: a slightly lower $S_z$ value results in lower clustering amplitude, which is compensated for by a higher linear bias. The results are also compared to those of the LRG HOD study in the DESI 1 per cent survey (S. Yuan et al. 2023). The linear bias reported for LRGs ranges from 1.94 at $z = 0.5$ to 2.31 at $z = 0.95$. While these values align with those of the CRS-LRG sample, the latter tends to exhibit lower bias at high redshifts. This is consistent with the target selection strategy, as the CRS-LRG sample selects brighter objects than the DESI sample. This can also be seen on Fig. 11, where at high redshift, the projected clustering of the CRS-LRG selection has a higher amplitude on large scales compared to the DESI-LRG sample, indicating a higher linear bias.

[6]https://github.com/keurfonluu/stochopy

[7]https://cosmoprimo.readthedocs.io/

**Table 7.** Results from the HOD minimisation for the six redshift bins, the 2 different photometric regions DECaLS and DES, and for the full sample labelled'ALL'. Columns are redshift bins, regions, the 5 fitted HOD parameters $\log_{10}(M_0)$, $\log_{10}(M_1)$, $\alpha$, $\log_{10}(M_{\rm cen})$ and $\sigma_M$, the applied velocity smearing $v_{smear}$ in km/$s$, the corresponding scaling factor for photo-$z$ error, the linear bias $b_{\rm lin}$, the satellite fraction $f_{\rm sat}$ and the reduced $\chi^2$ value of the best fit HOD model. Two values of $\chi^2_{\rm red}$ are reported, one using the full covariance and only the diagonal errors in parentheses. Masses are expressed in $M_\odot/h$.

| $z$ bins | Region | $\log_{10}(M_0)$ | $\log_{10}(M_1)$ | $\alpha$ | $\log_{10} M_{\rm cent}$ | $\sigma_M$ | $v_{smear}$ | $S_z$ | $b_{\rm lin}$ | $\log_{10}(\overline{M_h})$ | $f_{\rm sat}$ | $\chi^2_{\rm red}$ |
|---|---|---|---|---|---|---|---|---|---|---|---|---|
| $0.4 < z < 0.5$ | ALL | 12.68 | 13.94 | 1.11 | 12.76 | 0.22 | 10364 | 0.7 | 1.77 | 13.12 | 0.12 | 3.95 (*2.23*) |
| | DECaLS | 12.64 | 14.10 | 1.10 | 12.91 | 0.36 | | | 1.77 | 13.15 | 0.10 | 1.84 (*1.18*) |
| | DES | 12.83 | 13.89 | 1.00 | 12.73 | 0.17 | | | 1.76 | 13.12 | 0.12 | 1.71 (*1.36*) |
| $0.5 < z < 0.6$ | ALL | 12.93 | 14.07 | 1.05 | 12.94 | 0.23 | 10306 | 0.7 | 2.08 | 13.23 | 0.10 | 3.74 (*1.89*) |
| | DECaLS | 12.89 | 14.11 | 1.03 | 12.96 | 0.33 | | | 2.02 | 13.19 | 0.10 | 1.03 (*1.18*) |
| | DES | 12.71 | 14.20 | 1.03 | 12.92 | 0.14 | | | 2.08 | 13.25 | 0.10 | 2.48 (*1.41*) |
| $0.6 < z < 0.7$ | ALL | 12.78 | 13.78 | 1.02 | 12.73 | 0.24 | 10206 | 0.6 | 1.94 | 13.06 | 0.13 | 2.40 (*5.93*) |
| | DECaLS | 12.61 | 13.79 | 1.14 | 12.77 | 0.37 | | | 1.93 | 13.02 | 0.13 | 1.86 (*2.30*) |
| | DES | 12.85 | 14.05 | 0.92 | 12.84 | 0.24 | | | 2.02 | 13.13 | 0.10 | 2.56 (*2.05*) |
| $0.7 < z < 0.8$ | ALL | 12.55 | 14.23 | 1.05 | 12.93 | 0.09 | 11103 | 0.6 | 2.28 | 13.24 | 0.10 | 4.22 (*7.43*) |
| | DECaLS | 12.61 | 14.03 | 1.12 | 12.85 | 0.10 | | | 2.20 | 13.19 | 0.11 | 3.29 (*2.62*) |
| | DES | 12.79 | 14.11 | 0.99 | 12.91 | 0.17 | | | 2.27 | 13.21 | 0.10 | 2.86 (*1.14*) |
| $0.8 < z < 0.9$ | ALL | 12.52 | 13.91 | 1.03 | 12.80 | 0.20 | 13046 | 0.6 | 2.30 | 13.09 | 0.13 | 1.99 (*1.64*) |
| | DECaLS | 12.59 | 13.87 | 1.02 | 12.76 | 0.18 | | | 2.26 | 13.07 | 0.13 | 2.41 (*1.12*) |
| | DES | 12.67 | 13.92 | 1.06 | 12.82 | 0.15 | | | 2.37 | 13.13 | 0.12 | 0.96 (*0.70*) |
| $0.9 < z < 1.0$ | ALL | 12.41 | 14.21 | 0.98 | 12.91 | 0.21 | 17102 | 0.6 | 2.48 | 13.14 | 0.11 | 4.68 (*2.40*) |
| | DECaLS | 12.40 | 14.09 | 0.99 | 12.85 | 0.24 | | | 2.40 | 13.08 | 0.12 | 1.67 (*1.10*) |
| | DES | 12.54 | 14.36 | 0.94 | 12.96 | 0.17 | | | 2.63 | 13.21 | 0.10 | 2.22 (*1.69*) |

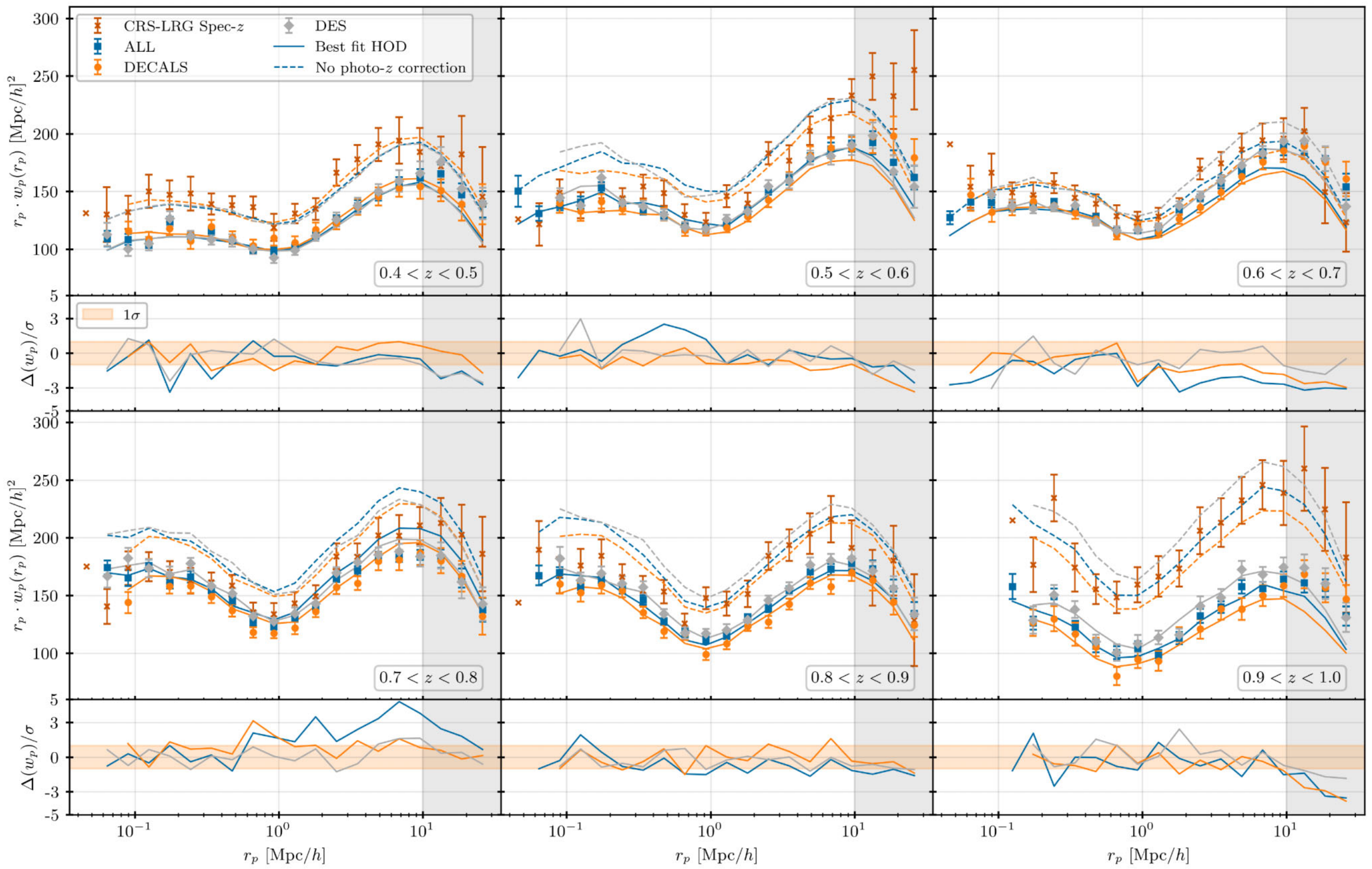

**Figure 9.** The projected correlation function multiplied by the transverse distance for the full CRS-LRG sample in blue (labelled as 'ALL') and the two different photometric regions, DES in grey, and DECaLS in orange. The dark orange points correspond to the CRS-LRG selected sample from the DESI DR1 data using spectroscopic redshift measurements. Each panel correspond to one redshift bin as labelled. The points are $w_p$ measurements with error bars obtained from 36 jackknife subregions. The solid lines are the best-fitting HOD results that include the smearing effect to account for photo-$z$ errors. The dashed lines are the corresponding clustering without smearing effect, i.e. showing the intrinsic clustering without photo-$z$ errors. The intrinsic clustering matches the CRS-LRG spec-$z$ sample, showing the impact of photo-$z$ errors. The smaller panels correspond to best-fitting residuals with photo-$z$ correction for each region, coloured accordingly. The orange shaded band represent the 1$\sigma$ region. The grey shaded area is not used for the HOD minimisation.

## 8 CONCLUSIONS

We have validated the 4MOST–CRS BG and LRG target catalogues selected from Legacy Surveys DR10.1 imaging using angular clustering, cross-correlations with DESI DR1, and (for LRGs) HOD modelling. These tests demonstrate that the adopted selections and veto masks deliver uniform, well-behaved clustering signals across the CRS footprint suitable for large-scale structure analyses.

For BG, applying the Legacy Survey `MASKBITS` {11, 12, 13} yields stable $w(\theta)$ measurements, with residual shifts $\Delta w(\theta)/\sigma(\theta)$ consistent with zero over the angular range of interest; stacked target–star maps support extending beyond the nominal `MASKBIT 1` radii around *Gaia* sources. For LRG, combining the same LS `MASKBITS` with the full set of unWISE W1 masks suppresses small-scale residuals and achieves convergence in $\Delta w(\theta)/\sigma(\theta)$.

A Limber-scaling test across BG $r$-band magnitude slices, using BE fits to $N(z)$ from DESI DR1 and a broken power-law description of $\xi(r)$ derived from $w_p(r_p)$, collapses the $w(\theta)$ curves to a near-common relation in both NGC (DECaLS) and SGC (DECaLS + DES). Post-scaling, cross-cap residuals lie within $\sim 1\sigma$, and the inferred horizontal and vertical offsets are essentially insensitive to the choice of clustering-evolution parameter ($\varepsilon = 0$ or $-1.3$). We interpret this as a uniformity check: the observed slice-to-slice differences are explained by $N(z)$ rather than spatially varying photometric systematics.

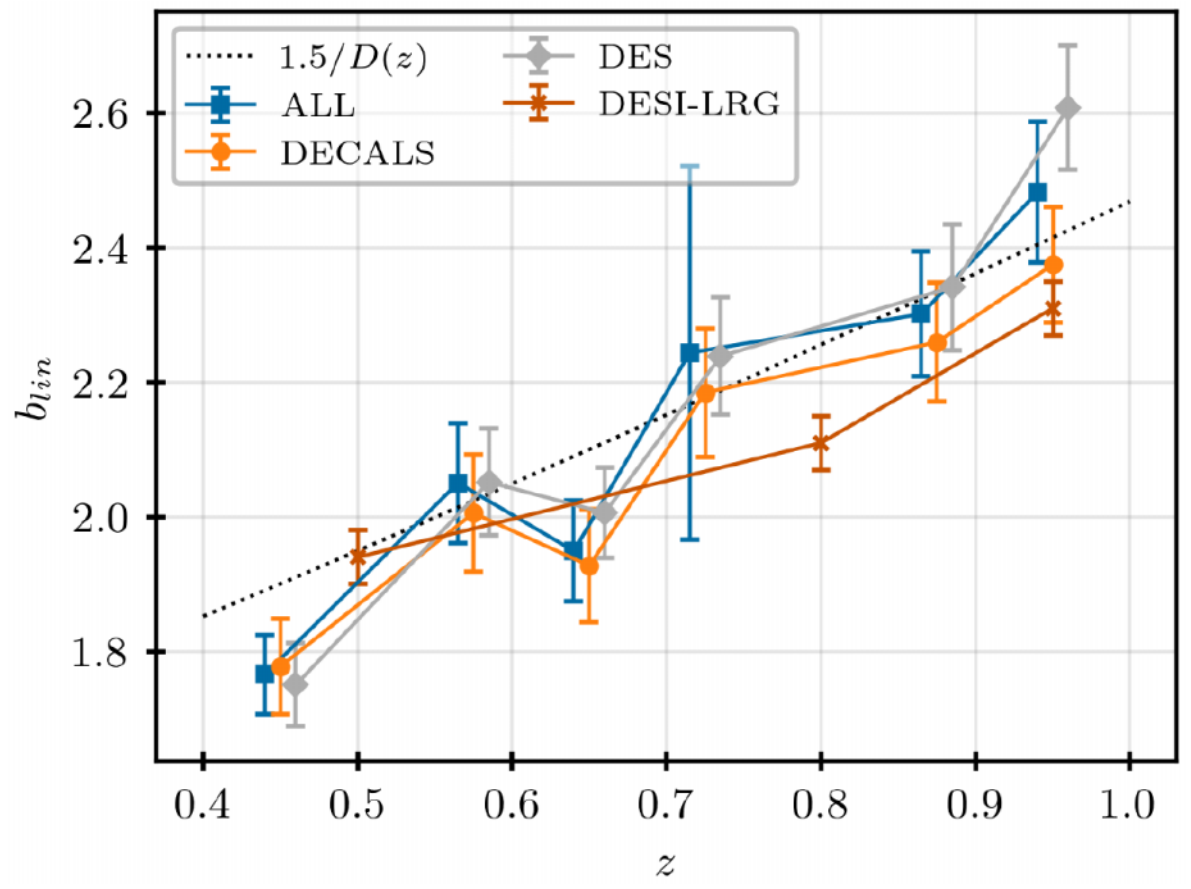

**Figure 10.** Evolution of the linear galaxy bias for the CRS-LRG sample for the two photometric regions DES (grey) and DECaLS (orange), and the full sample (ALL, blue). The points and the error bars are computed using the mean and the standard deviation from 50 HOD mocks spanning a range of HOD parameters around the best-fitting values reported in Table 7. Data points for each region have been shifted on the $x$-axis for clarity. The dark orange points with error bars come from the linear bias measured in the DESI 1 per cent survey S. Yuan et al. (2023). The dotted line is the predicted evolution of the inverse of the linear growth factor $D(z)$ in the baseline cosmology of our paper, representing the bias evolution for a constant clustering amplitude. The trend of the CRS-LRG sample is consistent with a constant clustering amplitude across time.

Clustering-based redshifts for BG reproduce the shape of the DESI DR1 $N(z)$ after normalisation in the fainter magnitude bins, with increased noise in the brightest bins. This independently supports the validity of the BG redshift distributions used in the Limber exercise and the robustness of the BG target selection.

For LRGs, the angular two-point function measured in photometric-redshift slices over $0.4 \leq z < 1.0$ is well described by a power law on $0.1^\circ < \theta < 0.8^\circ$ in both DECaLS and DES regions, with modest redshift evolution of slope and amplitude. One bin ($0.8 < z < 0.9$) shows elevated $\chi^2$ primarily at small scales, while large-angle differences between DECaLS and DES are within $\lesssim 1\sigma$; small-angle residuals reflect photometric-depth differences.

HOD fits to the LRG projected clustering provide a qualitative description consistent with recent LRG studies when photo-$z$ smearing is included, with satellite fractions of order $\sim 10$ per cent and linear-bias evolution consistent with expectations from the growth factor. Comparisons with a CRS-LRG-like spectroscopic selection from DESI DR1 behave as expected across redshift.

Taken together, these results show that the CRS BG and LRG target selections, together with the adopted masking, yield internally consistent clustering measurements across the survey area and validated redshift distributions for BG. This provides a sound basis for early CRS large-scale structure analyses and for cross-correlation work over the substantial overlaps with southern imaging surveys.

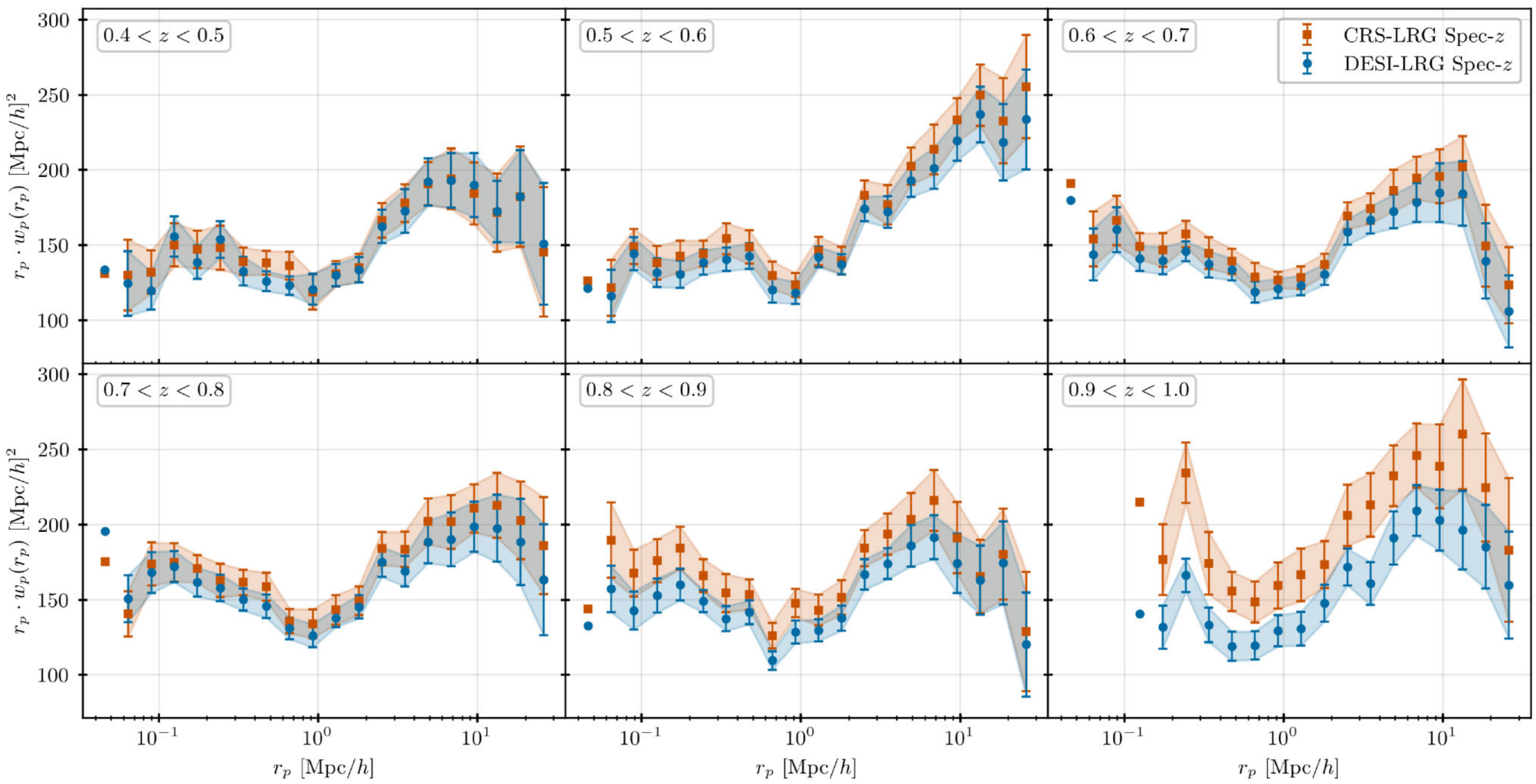

**Figure 11.** The projected correlation function multiplied by the transverse distance for spectroscopic samples from DESI DR1. The blue points with error bars correspond to a CRS-LRG-like sample selected from the DESI DR1 data, and the orange points correspond to the DESI LRG samples. Both samples come from the DESI DR1 footprint with `NTILE` $> 3$ to avoid regions with high fibre incompleteness. The shaded regions represent the size of the errors and are added for better visualisation. Error bars are obtained from 36 jackknife subregions. As expected from the CRS-LRG selections, the lowest redshift bins display similar amplitude to the DESI LRG sample and increase at higher redshifts.

## CONTRIBUTIONS

(i) Behnood Bandi: The major part of the analysis, BG clustering, cluster-z, maskings.
(ii) Antoine Rocher: BG selection and photometric systematics, HOD fits, LRG Angular Clustering in photo-z bins.
(iii) Aurélien Verdier: LRG target selection and forecasts.
(iv) Jon Loveday: Supervision, review and corrections.
(v) Zhuo Chen: Angular Clustering in photo-z bins and HOD fits.
(vi) Johan Richard: CRS management, masking tests, review and corrections
(vii) Jean-Paul Kneib: CRS management.
(viii) Tom Shanks and Michael Brown: Review and corrections

## ACKNOWLEDGEMENTS

Some parts of this work used the DiRAC Data Intensive service (CSD3) at the University of Cambridge, managed by the University of Cambridge University Information Services on behalf of the STFC DiRAC HPC Facility (https://www.dirac.ac.uk). The DiRAC component of CSD3 at Cambridge was funded by BEIS, UKRI, and STFC capital funding and STFC operations grants. DiRAC is part of the UKRI Digital Research Infrastructure. This work has made use of CosmoHub, developed by PIC (maintained by IFAE and CIEMAT) in collaboration with ICE-CSIC. It received funding from the Spanish government (grant EQC2021-007479-P funded by MCIN/AEI/10.13039/501100011033), the EU NextGeneration/PRTR (PRTR-C17.I1), and the Generalitat de Catalunya.

The DESI Legacy Imaging Surveys consist of three individual and complementary projects: the Dark Energy Camera Legacy Survey (DECaLS), the Beijing-Arizona Sky Survey (BASS), and the Mayall z-band Legacy Survey (MzLS). DECaLS, BASS, and MzLS together include data obtained, respectively, at the Blanco telescope, Cerro Tololo Inter-American Observatory, NSF's NOIRLab; the Bok telescope, Steward Observatory, University of Arizona; and the Mayall telescope, Kitt Peak National Observatory, NOIRLab. NOIRLab is operated by the Association of Universities for Research in Astronomy (AURA) under a cooperative agreement with the National Science Foundation. Pipeline processing and analyses of the data were supported by NOIRLab and the Lawrence Berkeley National Laboratory (LBNL). Legacy Surveys also uses data products from the Near-Earth Object Wide-field Infrared Survey Explorer (NEOWISE), a project of the Jet Propulsion Laboratory/California Institute of Technology, funded by the National Aeronautics and Space Administration. Legacy Surveys was supported by: the Director, Office of Science, Office of High Energy Physics of the U.S. Department of Energy; the National Energy Research Scientific Computing Center, a DOE Office of Science User Facility; the U.S. National Science Foundation, Division of Astronomical Sciences; the National Astronomical Observatories of China, the Chinese Academy of Sciences and the Chinese National Natural Science Foundation. LBNL is managed by the Regents of the University of California under contract to the U.S. Department of Energy. The complete acknowledgements can be found at https://www.legacysurvey.org/acknowledgment/.

This research used data obtained with the Dark Energy Spectroscopic Instrument (DESI). DESI construction and operations are managed by the Lawrence Berkeley National Laboratory. This material is based upon work supported by the U.S. Department of Energy, Office of Science, Office of High-Energy Physics, under Contract No. DE-AC02-05CH11231, and by the National Energy Research Scientific Computing Centre, a DOE Office of Science User Facility under the same contract. Additional support for DESI was provided by the U.S. National Science Foundation (NSF), Division of Astronomical Sciences under Contract No. AST-0950945 to the NSF's National Optical-Infrared Astronomy Research Laboratory; the Science and Technology Facilities Council of the United Kingdom; the Gordon and Betty Moore Foundation; the Heising-Simons Foundation; the French Alternative Energies and Atomic Energy Commission (CEA); the National Council of Humanities, Science and Technology of Mexico (CONAHCYT); the Ministry of Science and Innovation of Spain (MICINN), and by the DESI Member Institutions: www.desi.lbl.gov/collaborating-institutions. The DESI collaboration is honoured to be permitted to conduct scientific research on I'oligam Du'ag (Kitt Peak), a mountain with particular significance to the Tohono O'odham Nation. Any opinions, findings, and conclusions or recommendations expressed in this material are those of the author(s) and do not necessarily reflect the views of the U.S. National Science Foundation, the U.S. Department of Energy, or any of the listed funding agencies.

We acknowledge financial support from 'Action thématique de Cosmologie and Galaxies' (ATCG), funded by CNRS/INSU-IN2P3-INP, CEA and CNES, France. This work has made use of CosmoHub, developed by PIC (maintained by IFAE and CIEMAT) in collaboration with ICE-CSIC. It received funding from the Spanish government (grant EQC2021-007479-P funded by MCIN/AEI/10.13039/501100011033), the EU NextGeneration/PRTR (PRTR-C17.I1), and the Generalitat de Catalunya. We would also like to thank Boudewijn F. Roukema for their proofreading, comments, and suggestions and Hossein Zarei for their useful discussions.

In this work, we made use of ASTROPY (Astropy Collaboration 2022, 2025), NUMPY (C. R. Harris et al. 2020), PANDAS (P. D. Team 2020), MATPLOTLIB (J. D. Hunter 2007), HEALPY (K. M. Górski et al. 2005; A. Zonca et al. 2019), TREECORR (M. Jarvis et al. 2004), CORRFUNC (M. Sinha & L. H. Garrison 2020), ABACUSSUMMIT (N. A. Maksimova et al. 2021), KCORRECT (M. R. Blanton & S. Roweis 2007), and HODDIES (A. Rocher et al. 2023b) Python packages and TOPCAT (M. B. Taylor 2005).

## DATA AVAILABILITY AND CODE

The target catalogues were derived from the publicly available DESI Legacy Surveys DR10.1 imaging (https://www.legacysurvey.org/dr10/files/) and were selected following the target-selection criteria introduced by A. Verdier et al. (2025). The DESI DR1 spectroscopic data sets used for validation are available at https://data.desi.lbl.gov/doc/releases/dr1/. The Python scripts used to produce the results in Sections 3 to 5 are available at https://github.com/BehnoodBandi/crs_clustering.

## REFERENCES

Abazajian K. N. et al., 2009, ApJS, 182, 543
Astropy Collaboration, 2022, ApJ, 935, 167
Astropy Collaboration, 2025, Astropy, Zenodo. Available at: https://doi.org/10.5281/zenodo.14827951
Baugh C. M., Efstathiou G., 1993, MNRAS, 265, 145
Blanton M. R., Roweis S., 2007, AJ, 133, 734
Chaussidon E. et al., 2021, MNRAS, 509, 3904

Coil A. L., 2012, in Oswalt T.D., Keel W.C., eds, Planets, Stars and Stellar Systems, Springer, Dordrecht, p. 387
Davis M., Peebles P. J. E., 1983, ApJ, 267, 465
DESI Collaboration, 2025, AJ, preprint (arXiv:2503.14745)
de Jong R. S. et al., 2019, The Messenger, 175, 3
Dewdney P., Hall P., Schilizzi R., Lazio T., 2009, Proc. IEEE, 97, 1482
Dey A. et al., 2016, in Evans C. J., Simard L., Takami H. eds, Proc. SPIE Conf. Ser. Vol. 9908, Ground-Based and Airborne Instrumentation for Astronomy VI. SPIE, Bellingham, p. 99082C
Dey A. et al., 2019, AJ, 157, 168
Doumerg W. D. et al., 2025, A&A, 702, A155
Duncan K. J., 2022, MNRAS, 512, 3662
Edge A., Sutherland W., Kuijken K., Driver S., McMahon R., Eales S., Emerson J. P., 2013, The Messenger, 154, 32
Efstathiou G., Bernstein G., Tyson J. A., Katz N., Guhathakurta P., 1991, ApJ, 380, L47
Euclid Collaboration, 2025, A&A, 697, A1
Farrow D. J. et al., 2015, MNRAS, 454, 2120
Flaugher B. et al., 2015, AJ, 150, 150
Gatti M. et al., 2021, MNRAS, 510, 1223
Górski K. M., Hivon E., Banday A. J., Wandelt B. D., Hansen F. K., Reinecke M., Bartelmann M., 2005, ApJ, 622, 759
Groth E. J., Peebles P. J. E., 1977, ApJ, 217, 385
Hadzhiyska B., Eisenstein D., Bose S., Garrison L. H., Maksimova N., 2021, MNRAS, 509, 501
Hahn C. et al., 2023, AJ, 165, 253
Harris C. R. et al., 2020, Nature, 585, 357
Hildebrandt H. et al., 2021, A&A, 647, A124
Hunter J. D., 2007, Comput. Sci. Eng., 9, 90
Ivezić Ž. et al., 2019, ApJ, 873, 111
Jarvis M., Bernstein G., Jain B., 2004, MNRAS, 352, 338
Karademir G. S. et al., 2021, MNRAS, 509, 5467
Karademir G. S., Taylor E. N., Blake C., Cluver M. E., Jarrett T. H., Triani D. P., 2023, MNRAS, 522, 3693
Landy S. D., Szalay A. S., 1993, ApJ, 412, 64
Lesgourgues J., 2011, preprint (arXiv:1104.2932)
Limber D. N., 1953, ApJ, 117, 134
Liske J. et al., 2015, MNRAS, 452, 2087
Loveday J. et al., 2018, MNRAS, 474, 3435
Maddox S. J., Efstathiou G., Sutherland W. J., 1996, MNRAS, 283, 1227
Maksimova N. A., Garrison L. H., Eisenstein D. J., Hadzhiyska B., Bose S., Satterthwaite T. P., 2021, MNRAS, 508, 4017
Ménard B., Scranton R., Schmidt S., Morrison C., Jeong D., Budavari T., Rahman M., 2014, preprint (arXiv:1303.4722)
Morrison C. B., Hildebrandt H., Schmidt S. J., Baldry I. K., Bilicki M., Choi A., Erben T., Schneider P., 2017, MNRAS, 467, 3576
Myers A. D. et al., 2023, AJ, 165, 50
Naidoo K. et al., 2023, A&A, 670, A149
Navarro J. F., Frenk C. S., White S. D. M., 1997, ApJ, 490, 493
Newman J. A., 2008, ApJ, 684, 88
Peebles, 1980, The Large-Scale Structure of the Universe. Princeton Univ. Press, Princeton, NJ
Phillipps S., Fong R., Ellis R. S., Fall S. M., MacGillivray H. T., 1978, MNRAS, 182, 673
Planck Collaboration VI, 2020, A&A, 641, A6
Richard J. et al., 2019, The Messenger, 175, 50
Rocher A. et al., J. Cosmol. Astropar. Phys., 2023, 016
Rocher A., Ruhlmann-Kleider V., Burtin E., De Mattia A., 2023b, J. Cosmol. Astropart. Phys., 2023, 033
Ross A. J. et al., 2024, J. Cosmol. Astropart. Phys., 2025, 125
Sawangwit U., Shanks T., Abdalla F. B., Cannon R. D., Croom S. M., Edge A. C., Ross N. P., Wake D. A., 2011, MNRAS, 416, 3033
Sinha M., Garrison L., 2019a, in Majumdar A., Arora R., eds, Software Challenges to Exascale Computing. Springer Singapore, Singapore, p. 3, https://doi.org/10.1007/978-981-13-7729-7_1
Sinha M., Garrison L., 2019b, in Majumdar A., Arora R., eds, Vol. 964, Software Challenges to Exascale Computing. Springer Singapore, Singapore, p. 3,
Sinha M., Garrison L. H., 2020, MNRAS, 491, 3022
Taylor M. B., 2005, in Shopbell P., Britton M., Ebert R., eds, ASP Conf. Ser. Vol. 347, Astronomical Data Analysis Software and Systems XIV. Astron. Soc. Pac., San Francisco, p. 29
Team P. D., 2020, pandas-dev/pandas: Pandas, Zenodo. Available at : https://doi.org/10.5281/zenodo.3509134
Tempel E. et al., 2025, A&A, 703, A261
Valdes F., Gruendl R., DES Project, 2014, in Manset N., Forshay P., eds, ASP Conf. Ser. Vol. 485, Active Galaxies. Astronomical Data Analysis Software and Systems XXIII. Astron. Soc. Pac., San Francisco, p. 379
Verdier A. et al., 2025, MNRAS, 545, staf2116 (VR25)
Wechsler R. H., Tinker J. L., 2018, ARA&A, 56, 435
Wright A. H. et al., 2024, A&A, 686, A170
Wright E. L. et al., 2010, AJ, 140, 1868
Wu C.-F. J., 1986, Ann. Stat., 14, 1261
Yuan S., Garrison L. H., Hadzhiyska B., Bose S., Eisenstein D. J., 2022, MNRAS, 510, 3301
Yuan S. et al., 2023, MNRAS, 530, 947
Zarrouk P. et al., 2021, MNRAS, 509, 1478
Zhai Z. et al., 2017, ApJ, 848, 76
Zheng Z. et al., 2005, ApJ, 633, 791
Zhou R. et al., 2021, MNRAS, 501, 3309
Zhou R. et al., 2023, AJ, 165, 58
Zonca A., Singer L., Lenz D., Reinecke M., Rosset C., Hivon E., Gorski K., 2019, J. Open Source Softw., 4, 1298
Zou H. et al., 2017, PASP, 129, 064101

This paper has been typeset from a TEX/LATEX file prepared by the author.